# Zero and Finite Temperature Phase Diagram of the Spinless Fermion Model in Infinite Dimensions

G.S. Uhrig[1]

*Institut für Theoretische Physik, Universität zu Köln,*

*D-50937 Köln, Germany.*

R. Vlaming

*Instituut voor theoretische fysica, Universiteit van Amsterdam,*

*NL-1018 XE, The Netherlands.*

## Abstract

The phase diagram of the model of spinless fermions with repulsive nearest neighbour interaction is calculated analytically on a hypercubic lattice in infinite dimensions ($d \to \infty$). In spite of its simplicity the model displays a rich phase diagram depending on the doping $\delta$, the interaction $U$ and the temperature $T$. The system can be in the homogeneous phase (HOM), the nonsegregated AB charge density wave (AB-CDW), the AB phase separation region (PS-AB/HOM; coexistence of AB-CDW and HOM), the incommensurate phase (IP) or the IP phase separation region (PS-AB/IP; coexistence of AB-CDW and IP). We identify three important values of the interaction $U_{\text{IPL}} = 0.572 < U_{\text{IPH}} = 1.914 < U_{\text{IP/PS}} = 4.212$ which distinguish four intervals of $U$. These imply four different types of phase diagrams. In all the three phase diagrams with $U$ below $U_{\text{IP/PS}}$ the IP appears. We propose a *new* general ansatz for the order parameter of this phase. A competition between the IP, the PS-AB/IP and the PS-AB/HOM is found. The relevance of our findings for the phase scenario of the Hubbard model is shown.

---

[1]Present address: Laboratoire de Physique des Solides, Université Paris-Sud, bât. 510, Centre Universitaire, F-91405 ORSAY Cédex

# 1 Introduction

The interest in non-perturbative results for highly correlated itinerant fermion systems revived during the last years. Although in general analytically exact solutions of fermionic models with strong interaction are beyond reach, particular cases in the parameter space are solvable. The dimension plays an important role in this respect. In dimension $d = 1$, for example, the Bethe ansatz is at our disposal by which numerous models are analytically tractable.

The other extreme, i.e. $d \to \infty$, is a useful tool for classical systems and has been extended to itinerant quantum systems about six years ago [1, 2]. This technique is successful for a variety of itinerant quantum models such as the Hubbard model [3–8], the $t$-$J$ model [9, 10], the Anderson model of disorder [11], the periodic Anderson model [12] and the Falicov-Kimball model [13, 14]. An extensive review is written by Vollhardt [15].

In the present paper we study the model of spinless fermions with nearest neighbour interaction on a hypercubic lattice in the limit of infinite dimensions. It is our goal to find the phase diagram in the ground state as well as for finite temperatures. Our analysis shows a competition between the homogeneous phase (HOM), the nonsegregated AB charge density wave (AB-CDW), the AB phase separation region (PS-AB/HOM), the incommensurate phase (IP) and the IP phase separation region (PS-AB/IP). Phase separation means the segregation of phases: one part of the system has a doping above the average, another has a doping below the average. In the case PS-AB/HOM, a hole-rich homogeneous phase and a hole-poor AB charge density wave coexist. In the case PS-AB/IP, a hole-rich incommensurate phase and a hole-poor AB charge density wave coexist. These phenomena are for the first time comprised in a single model simple enough to be analysed analytically.

Some of the phases mentioned above attract attention since the discovery of the high-$T_c$ superconducting materials. Phase separation is a manifestation of an attractive tendency. This is found for example in the $t$-$J$ model [16–19] and in effective Heisenberg models [20]. Experimentally this is observed in lanthanum cuprates [21]. It is suggested that bound states of two or more particles will occur close to the region where the phases separate [16, 17, 19]. Spontaneous symmetry breaking with incommensurate order parameter was found theoretically in itinerant quantum antiferromagnets [22] for the Hubbard model [23–27], the $t$-$J$ model [28–30], and the Falicov-Kimball model [31]. Experimentally it was observed that the incommensurate order parameter couples to the superconducting order parameter [32].

The model of interacting spinless fermions can serve as a relatively simple model which allows to test certain approaches like the limit of infinite dimensions. Moreover it allows to obtain an overview over the possible phase scenarios in more elaborate models and the competition between these scenarios. As we will see in the course of this paper many qualitative features of the infinite dimensional spinless fermion model are encountered in the infinite dimensional Hubbard model and other models, too.

Furthermore there are physical systems which can be described with the model of interacting spinless fermions such as strongly polarised systems, e.g. a lattice gas description of a $^3$He in a strong magnetic field, or ferro (ferri) magnetic electronic systems where one spin-band is filled. In these systems only electrons of one spin species are itinerant. The latter situation is realised in magnetite ($Fe_3O_4$), where the lowest spin-up band is half filled, leading to a metallic conductivity above a temperature $T_v \simeq 119$K at atmospheric pressure. At $T_v$ the system undergoes the Verwey-transition into an insulator [33]. Cullen and Callen [34, 35] first suggested a model of spinless fermions with nearest neighbour interaction to describe this transition.

The model itself can be solved exactly in $d = 1$ by Bethe ansatz [36, 37]. At half filling the solution describes a transition from a homogeneous phase to a AB charge density ordered phase at a finite value of the interaction. The transition is discussed in detail by Shankar [38], who also calculates various response functions. A transition to a charge density ordered state at a *finite* interaction turns out to be a peculiarity of one dimension [39]. For $d > 1$ at half filling on a bipartite lattice the transition is shifted to arbitrarily small interactions according to renormalisation group results [39, 40]. Using a quantum Monte Carlo simulation technique the problem was studied in two dimensions by Gubernatis *et al.* [41]. For finite temperatures and half filling in $d > 1$ Lemberger and Macris [42] proved rigorously the presence of an AB-CDW at sufficiently large values of the interaction. Apart from Yang and Yang [37], who include certain aspects of finite doping, the research so far concentrates on the half filled case.

It is the objective of the present paper to provide an exhaustive description of the phase diagram of repulsively interacting spinless fermions thereby extending and compiling previous results [43, 44]. The striking analogy between certain results concerning the incommensurate phases in the model of spinless fermions and in the extended Hubbard model [45] triggers the interest in a circumstantial discussion of such calculations.

The paper is organised as follows. In sect. 2 the Hamiltonian and the relevant results for $d \to \infty$ are given. Subsequently in sect. 3, the equations are derived which determine the second and the first order transition lines at zero temperature. From these the $T = 0$ phase diagram is constructed. The phase diagram for finite temperatures is constructed in sect. 4. In sect. 5 we conclude our work by a summary and a discussion of the implications of our results for the understanding of the phase scenario in the (extended) Hubbard model.

## 2  Hamiltonian and limit of infinite dimensions

We study the model of itinerant spinless fermions with nearest neighbour hopping $t$ and nearest neighbour repulsion $U$ on a hypercubic lattice in $d \to \infty$ dimensions. The Hamiltonian

is

$$\hat{H} := \sum_{i,j} t_{ij}\, \hat{c}_i^+ \hat{c}_j + \tfrac{1}{2} \sum_{i,j} U_{ij}\, \hat{n}_i\, \hat{n}_j - \sum_i \mu\, \hat{n}_i\, , \tag{1}$$

where $\hat{c}_i^+$ ($\hat{c}_i$) are the creation (destruction) operators for fermions on site $i$, $\hat{n}_i := \hat{c}_i^+ \hat{c}_i$ and $U_{ij}$, $t_{ij}$ are zero if $i, j$ are not nearest neighbours. The hopping matrix elements are scaled as $t_{ij} := -t/\sqrt{2d}$ and the interaction matrix elements as $U_{ij} = U/(2d)$, given that the sites $i$ and $j$ are nearest neighbours. This scaling guarantees that the model remains non-trivial in the limit $d \to \infty$ [1, 2]. Then the dispersion relation $\epsilon_{\vec{k}}$ is

$$\epsilon_{\vec{k}} = -\frac{2t}{\sqrt{2d}} \sum_{n=1}^{d} \cos(k_n)\, . \tag{2}$$

Since the hypercubic lattice is bipartite, the Hamiltonian is particle-hole symmetric and we may restrict the filling $n$ to $0 < n \leq 1/2$ or, equivalently, the doping $\delta$ to $\delta := 1/2 - n \geq 0$. Throughout the paper we will work in units of $\hbar = 1$, $k_{\rm B} = 1$ and $t = 1$. The lattice spacing is set to unity.

It is known that in the limit of large dimensions the self-energy becomes site-diagonal and thus $\vec{k}$ independent [15]. For interactions as the one considered in (1) which are not on-site the Hartree approximation becomes exact [2]. We show that on the two-particle level the random phase approximation (RPA) becomes exact on $d \to \infty$. The proofs of these statements rely all on the resummation of the corresponding perturbation series on a diagrammatic level. In order to cover both the Hartree approximation for the one-particle quantities and the random phase approximation for the two-particle quantities we argue on the basis of the generating functional $\Phi[G]$ introduced by Baym and Kadanoff [46, 47]. Furthermore it is advantageous that this line of argument can easily be extended to $1/d$ corrections [48].

The perturbation diagrams of which the sum equals the generating functional $\Phi[G]$ are classified according to their contribution in powers of $1/d$. In fig. 1 three examples are given. The diagram in fig. 1(a) is of order $d^0$ whereas the diagrams in 1(b) and 1(c) are or order $d^{-1}$. The power counting for the diagrams relies on the proportionality of the propagator in real space $G_{ij} \propto d^{-\|\vec{r}_i - \vec{r}_j\|/2}$ [1, 2] where the Manhattan metric $\|\cdot\|$ counts the minimum number of hops to reach $\vec{r}_j$ starting from $\vec{r}_i$. The proportionality can easily be derived by expanding the propagator in the scaled hopping. For instance, in the diagram in fig. 1(a) both propagators are local ($i = j$). The factor $1/d$ from the scaling of the interaction is compensated by the $2d$ realisations of this diagram on the lattice (there are $2d$ nearest neighbours). In the diagram in fig. 1(b) both propagators go from one site to one of its nearest neighbours ($\|\vec{r}_i - \vec{r}_j\| = 1$) which adds an extra factor $1/d$.

The one- and the two-particle irreducible vertices are found from $\Phi[G]$ by functional derivation with respect to the Green function of the interacting system $\Sigma = \delta\Phi/\delta G$ and $\Xi = \delta^2\Phi/\delta G \delta G'$. Inserting $\Sigma$ ($\Xi$) into the Dyson equation (the Bethe-Salpeter equation) yields the full Green function (the general two-particle correlation function). The ensuing diagrammatic series are just the self-consistent Hartree and random phase approximations. They are depicted in figs. 2(a) and 2(b).

Two further consequences of $d \to \infty$ are important in our calculations. First, the DOS of the non-interacting system is a Gaussian

$$\rho(\epsilon) := \frac{1}{N} \sum_{\vec{k}} \delta(\epsilon - \epsilon_{\vec{k}}) = \frac{1}{\sqrt{2\pi}} \exp\left(-\frac{\epsilon^2}{2}\right) , \tag{3}$$

where $N$ is the system size. Eq. (3) follows from the central limit theorem since the dispersion $\epsilon_{\vec{k}}$ in (2) is a sum of terms each depending only on one component of $\vec{k}$ [1]. Second, the two-energy density $\rho(\epsilon_1, \epsilon_2, \vec{k})$ which occurs in the calculation of response functions (see below) is a Gaussian, too. It depends on $\vec{k}$ only via the parameter $\eta_{\vec{k}}$ [2, 49]

$$\rho(\epsilon_1, \epsilon_2, \eta_{\vec{k}}) = \frac{1}{N^2} \sum_{\vec{k}'} \sum_{\vec{k}''} \delta(\epsilon_1 - \epsilon_{\vec{k}'}) \delta(\epsilon_2 - \epsilon_{\vec{k}''}) \delta(\vec{k} - (\vec{k}' - \vec{k}'')) =$$

$$= \frac{1}{2\pi\sqrt{1 - \eta_{\vec{k}}^2}} \exp\left(-\frac{\epsilon_1^2 - 2\epsilon_1 \epsilon_2 \eta_{\vec{k}} + \epsilon_2^2}{2(1 - \eta_{\vec{k}}^2)}\right) \quad \text{with} \tag{4a}$$

$$\eta_{\vec{k}} := \frac{1}{d} \sum_{n=1}^{d} \cos(k_n) . \tag{4b}$$

In all dimensions $-1 \leq \eta_{\vec{k}} \leq 1$ holds. We will call $\eta_{\vec{k}}$ the parameter of incommensurability since the commensurate values of the wave vector $\vec{k}$ which fulfil $\eta_{\vec{k}} = \eta_0$ for $-1 < \eta_0 < 1$ are of measure zero in the Brillouin zone. For $d \to \infty$ $\eta_{\vec{k}}$ is zero for most possible values of $\vec{k}$ as can be seen comparing the scaling of $\eta_{\vec{k}}$ with the scaling of $\epsilon_{\vec{k}}$ in (2). But for certain values of $\vec{k}$ which are of measure zero in the Brillouin zone for $d \to \infty$ $\eta_{\vec{k}}$ does not vanish. These values are physically very important since the value of $\eta$ is fixed externally; it is impressed on the system from the outside. For instance, one has $\eta_{\vec{Q}} = -1$ for the nesting vector $\vec{Q} := (\pi, \pi, \ldots, \pi)^{\dagger}$. In contrast to this, the dispersion $\epsilon_{\vec{k}}$ is involved in integrated quantities. This explains why both quantities $\eta_{\vec{k}}$ and $\epsilon_{\vec{k}}$ are physically relevant in spite of their different scaling in $1/d$.

## 3  Phase diagram at zero temperature

Two regions of the phase diagram are known. At half filling and large interaction the AB-CDW is present and for very low values of the interaction and large doping the system is in the homogeneous phase. In the first case, the interaction requires that there are no two particles adjacent to each other. At half filling this requirement can only be met in the AB-CDW. In the latter limit, the interaction becomes finally irrelevant. So the system is in the HOM phase.

Starting in the HOM phase for low values of the interaction and for low filling one approaches the transition boundaries to other phases by continuously increasing the interaction and the filling. At the boundary in the $U - \delta$ plane at which a second order transition from the HOM phase into another phase occurs the system becomes infinitely susceptible to density fluctuations characterised by a certain wave vector $\vec{k}$. The divergence of the density-density response function $\chi(\vec{k})$ indicates the instability of the HOM phase. This stability analysis is carried out in the following first subsection. The resulting second order boundary marks the maximum region the HOM phase may occupy in the $U - \delta$ plane.

Second order transitions can be superseded by first order transitions. The boundary which marks a first order transition between two phases can be constructed only if the ground state energies of both phases are known. In the second subsection we will construct the first order transition line between HOM and AB-CDW as well as the the first order transition line between IP and AB-CDW

## 3.1 Second order transitions

In the HOM phase the Green function $G$ and the density-density response function $\chi$ are diagonal in $\vec{k}$-space. The Green function as function of the Matsubara frequency $\omega_n$ and the wave vector $\vec{k}$ is

$$G(i\omega_n, \vec{k}) = (i\omega_n - \epsilon_{\vec{k}} + \tilde{\mu})^{-1} . \tag{5}$$

The only trace of the interaction is the renormalised chemical potential

$$\tilde{\mu} = \mu - \Sigma_{\text{AVE}} = \mu - nU . \tag{6}$$

It equals the true chemical potential reduced by the space-averaged self-energy which is $nU$. Integrating (3) yields the relation between $\tilde{\mu}$ and doping $\delta$

$$\delta = -\frac{1}{2} \operatorname{erf}\left(\frac{\tilde{\mu}}{\sqrt{2}}\right) . \tag{7}$$

The RPA series for $\chi(\vec{k})$ derived from fig. 2(b) is shown in fig. 2(c). It is a geometric series with the factor $-U\eta_{\vec{k}}\chi_0(\vec{k})$. This follows from the evaluation rules for two-particle diagrams: the wavy interaction line with the wave vector $\vec{k}$ stands for the negative Fourier transform of $U_{ij}$ which is $-U\eta_{\vec{k}}$. A single bubble represents $\chi_0(\vec{k})$, i.e. one has

$$\chi(\vec{k}) = \frac{\chi_0(\vec{k})}{1 + U\eta_{\vec{k}}\chi_0(\vec{k})} . \tag{8}$$

The value of $\chi_0(\vec{k})$ is explicitly given by

$$\chi_0(\vec{k}) := -\frac{T}{N} \sum_{\omega_n, \vec{k}'} G(i\omega_n, \vec{k}' + \vec{k}) G(i\omega_n, \vec{k}') \implies \tag{9a}$$

$$\chi_0(\eta_{\vec{k}}) = -\int_{-\infty}^{\infty} d\epsilon_1 \int_{-\infty}^{\infty} d\epsilon_2 \, \rho(\epsilon_1, \epsilon_2; \eta_{\vec{k}}) T \sum_{\omega_n} \frac{(i\omega_n - \epsilon_1 + \tilde{\mu})^{-1} - (i\omega_n - \epsilon_2 + \tilde{\mu})^{-1}}{\epsilon_1 - \epsilon_2} \tag{9b}$$

$$= -\int_{-\infty}^{\infty} d\epsilon_1 \int_{-\infty}^{\infty} d\epsilon_2 \, \rho(\epsilon_1, \epsilon_2; \eta_{\vec{k}}) \frac{f_{\text{F}}(\epsilon_1) - f_{\text{F}}(\epsilon_2)}{\epsilon_1 - \epsilon_2} \tag{9c}$$

$$= \frac{1}{\sqrt{2\pi(1-\eta)}} \int_1^{2/(1+\eta)} \frac{dt}{t\sqrt{2 - (1+\eta)t}} \exp\left(\frac{-\tilde{\mu}^2}{2}t\right) . \tag{9d}$$

Partial fraction expansion leads to (9b) and the two-energy density (4) is used to transform the $\vec{k}$-sum over the Brillouin zone. The Fermi functions $f_F(\epsilon) := (1 + \exp(\beta(x - \tilde{\mu}))^{-1}$ in (9c) are step functions at $T = 0$ but at finite temperatures the formula holds in the form (9c).

The central result of (8) and (9) is that $\chi$ depends on $\vec{k}$ only via the parameter $\eta_{\vec{k}}$ [43].

The second order boundaries are found from

$$0 = \chi(\eta_{\vec{k}})^{-1} = \chi_0(\eta_{\vec{k}})^{-1} + U\eta_{\vec{k}} \ . \tag{10}$$

The case $\eta = -1$ corresponds to $\vec{k} = \vec{Q}$. Its divergence indicates the transition to the AB-CDW. The integral (9d) simplifies for $\eta = -1$ and one has

$$\frac{1}{U} = \frac{-1}{2\sqrt{2\pi}} \text{Ei}(-\tilde{\mu}^2/2) \tag{11}$$

for the corresponding values of the interaction. Henceforth, we will drop the explicit notation of the $\vec{k}$-dependence of $\eta$.

The analysis of the relation (10) for general values of $\eta$ is similar to the one of Lindner and Schumacher [23] for the two dimensional Hubbard model and to the one of Freericks [31] for the infinite dimensional Falicov-Kimball model. Coming from low interaction and low filling we look for the minimum value of the interaction at which the susceptibility diverges for some value of $\eta$. We call the value of the parameter of incommensurability at which this happens $\eta_{\text{opt}}$ since it is the most favourable (optimum) value of $\eta$ for a phase transition.

The results are depicted in fig. 3. The dotted curve in fig. 3(a) displays the line at which $\chi(-1)$ diverges. It is found from (7) and (11). The solid curve in fig. 3(a) displays the line at which $\chi(\eta_{\text{opt}})$ diverges. It is found from (7), (10) and the derivative of (10) with respect to $\eta$. For the latter curve the dependence of $\eta_{\text{opt}}$ on the doping is given in fig. 3(b). We observe that $\eta_{\text{opt}}(\delta)$ is continuous for $\delta \to 0$ but jumps to $-1$ at $\delta = \delta_{\text{END}}$ where the two curves in fig. 3(a) intersect. The essential point is that for $0 < \delta \leq \delta_{\text{END}}$ the system is more susceptible towards an incommensurate CDW than towards the usual AB-CDW [43].

The optimum value $\eta_{\text{opt}}$ does not exceed $\eta_{\text{opt}} \lesssim -0.793$ at zero temperature. This is directly related to the nature of the interaction. Since the repulsive interaction takes place between particles on nearest neighbour sites an AB-structure is favoured even by the interaction alone. On the contrary, a local interaction alone does not favour any specific spatial pattern. Therefore one may expect higher values of $\eta_{\text{opt}}$ in models with local interaction. Indeed, Freericks [31] finds values of $\eta_{\text{opt}}$ up to one, i.e. $-1 \leq \eta_{\text{opt}} \leq 1$, for the Falicov-Kimball model. For the Hubbard model in $d = 3$ Lindner and Schumacher [23] find $-1 \leq \eta_{\text{opt}} \leq 0$, just as Doll, Dzierzawa, Frésard and Wölfle [25–27] do. On the other hand, there are indications that the quantum fluctuations in the Hubbard model prevent $\eta_{\text{opt}}$ from deviating too far from $-1$ [50].

The vicinity of the perfect nesting singularity at $\delta = 0$ is of particular interest. In the limit $\delta \to 0$ the optimum value of the parameter of incommensurability $\eta$ is related to the

doping by

$$1 + \eta_{\mathrm{opt}} = \frac{2\pi}{s^2}\delta^2 + \mathcal{O}(\delta^3) \quad \text{where} \tag{12a}$$

$$2s\mathrm{D}(s) = 1 \Rightarrow s \simeq 0.924139 \, . \tag{12b}$$

The function $\mathrm{D}(x)$ is the Dawson function defined by $\mathrm{D}(x) := -\frac{i}{2}\sqrt{\pi}\exp(-x^2)\operatorname{erf}(ix)$. With (12) the relation between $\delta$ and $U$ in the limit $\delta \to 0$ or equivalently in the limit $U \to 0$ can be derived from (10) and (7)

$$\delta_{\mathrm{IP}} = \frac{1}{\sqrt{\pi}}\exp\left(-\sqrt{2\pi}/U - C/2 + S\right) \, , \tag{13a}$$

$$S = \ln(s) - \frac{1}{2\sqrt{\pi}}\int_{-\infty}^{\infty} dt \exp(-(t-s)^2) \ln(t^2) \simeq 0.24536 \, , \tag{13b}$$

where $C$ is Eulers constant: $C \simeq 0.57722$. The corresponding relation for fixed value $\eta = -1$ is derived from (11) and (7)

$$\delta_{\mathrm{AB}} = \frac{1}{\sqrt{\pi}}\exp\left(-\sqrt{2\pi}/U - C/2\right) \, . \tag{14}$$

The fact that $\delta_{\mathrm{IP}} > \delta_{\mathrm{AB}}$ at fixed $U$ shows that the divergence of the susceptibility for an incommensurate value of the wave vector $\vec{k}$ occurs before the divergence for the nesting vector $\vec{Q}$.

Meanwhile van Dongen found that the relations (12), (13) and (14) are also relevant in the extended Hubbard model in the weak coupling limit [45]. The main difference to the situation in the model of spinless fermions consists in a renormalisation of the interaction due to fluctuations.

Using (12) and expanding (4b) for small deviations of $\vec{k}$ from $\vec{Q}$ we find

$$\frac{1}{\sqrt{d}}|\vec{k}-\vec{Q}| \approx \frac{2\sqrt{\pi}}{s}\delta \, . \tag{15}$$

The deviation between the optimum wave vector $\vec{k}$ and the nesting wave vector $\vec{Q}$ characterising the AB-CDW is proportional to the doping. Shraiman and Siggia found a similar proportionality for the incommensurate pitch of a spiral order in a quantum antiferromagnet [22].

## 3.2 First order transition HOM⟷AB-CDW

In this subsection we derive that the AB-CDW is stable only at half filling since it is unstable towards phase separation at finite doping. Furthermore we discuss a promising ansatz for the order parameter in the IP. First, however, we have to introduce the description of the AB-CDW.

In the AB-CDW particle densities on the two sublattices of a bipartite hypercubic lattice are different. Let us define $n$ to be the average particle density and $b$ to be the difference between the particle density on the B-sublattice and $n$

$$n := (\langle n_{a\in A}\rangle + \langle n_{b\in B}\rangle)/2 \tag{16a}$$
$$b := \langle n_{b\in B}\rangle - n . \tag{16b}$$

In Hartree theory the average self-energy is $nU$ which is taken into account by the renormalisation of the chemical potential (6). A finite value of $b$ induces an alternating field $\Delta \exp(i\vec{Q})$ ($\vec{Q} = (\pi, \pi, \ldots, \pi)^\dagger$ is the nesting vector) which is $\Delta = bU$ on the sites of sublattice A and $-\Delta$ on the sites of sublattice B. In this way the translational symmetry is broken and the Green function and the susceptibility are no longer diagonal in $\vec{k}$-space. The mode at $\vec{k}$ couples to the mode at $\vec{k} + \vec{Q}$. Since $\vec{k} + \vec{Q} + \vec{Q} = \vec{k}$ up to a reciprocal lattice vector it is sufficient to consider $2 \times 2$ matrices in the AB-CDW in contrast to the scalars in the HOM phase. The Green function becomes block diagonal in $\vec{k}$-space for finite $b$

$$\boldsymbol{G}(i\omega_n, \vec{k}) = \frac{1}{(i\omega_n + \tilde{\mu})^2 - \epsilon_{\vec{k}}^2 - \Delta^2} \begin{bmatrix} i\omega_n + \epsilon_{\vec{k}} + \tilde{\mu} & \Delta \\ \Delta & i\omega_n - \epsilon_{\vec{k}} + \tilde{\mu} \end{bmatrix} . \tag{17}$$

From $n = (T/N) \sum_{\vec{k}} \sum_n G_{1,1}(i\omega_n, \vec{k})$ and from $b = (T/N) \sum_{\vec{k}} \sum_n G_{1,2}(i\omega_n, \vec{k})$ one obtains

$$n = \int_{-\infty}^{\infty} \frac{-\omega \, d\omega}{\sqrt{\omega^2 - \Delta^2}} f_F(\omega) \rho(\sqrt{\omega^2 - \Delta^2}) \Theta(\omega^2 - \Delta^2) , \tag{18a}$$

$$U^{-1} = \int_{-\infty}^{\infty} \frac{d\omega}{\sqrt{\omega^2 - \Delta^2}} f_F(\omega) \rho(\sqrt{\omega^2 - \Delta^2}) \Theta(\omega^2 - \Delta^2) . \tag{18b}$$

Here $\Theta(x)$ is the Heaviside step function. These two equations determine $\tilde{\mu}$ and $\Delta$ ($b$, respectively) for externally given values for $n$ and $U$.

Evaluating the RPA series in fig. 2(c) with the propagator in (17) yields a $2 \times 2$ susceptibility $\boldsymbol{\chi}$. The instability of the AB-CDW is manifest in $\det(\boldsymbol{\chi}) < 0$ for all $\delta > 0$. At half filling the AB-CDW is stable. This instability can be traced back to the non-convexity of the ground state energy which leads eventually to a first order transition and phase separation.

The ground state energy $E(n, \Delta)$ can be found by integration from $\partial E(n, \Delta)/\partial n|_\Delta = \mu$ and $\partial E(n, \Delta)/\partial \Delta|_n = 0$ with the help of (18) or using the relation between the generating functional $\Phi[\boldsymbol{G}]$ and the free energy or the ground state energy, respectively [47], (see also the corresponding subsection for finite temperatures). The result is

$$E(n, \Delta; \eta) = \frac{\Delta^2}{2U} + \frac{U}{2}n^2 + \int_{-\infty}^{\mathrm{sgn}(\tilde{\mu})\sqrt{\tilde{\mu}^2 - \Delta^2}\Theta(\tilde{\mu}^2 - \Delta^2)} d\omega \, \mathrm{sgn}(\omega)\sqrt{\omega^2 + \Delta^2} \rho(\omega) . \tag{19}$$

The dashed-dotted curve in fig. 4 depicts a generic result for the ground state energy in the AB-CDW at $U = 1.2$. The dotted curve shows the continuation in the HOM phase for $\delta > \delta_{\text{div}-AB}$. The system will segregate into a spatial region with a hole-rich HOM phase and into a spatial region with the AB-CDW without holes. Theoretically one has to pass to the convex hull of the dashed-dotted curve in fig. 4 which is a straight line from the point at half filling ($\delta = 0$) and the point at $\delta = \delta_{\text{PS}-AB}$ (Maxwell construction). The points ($\delta_{\text{PS}-AB}, U$) result from

$$\delta_{\text{PS}-AB} = -\frac{1}{2}\,\text{erf}\left(\frac{\tilde{\mu}_{\text{PS}-AB}}{\sqrt{2}}\right)\,, \tag{20a}$$

$$\frac{1}{U} = \frac{1}{2\sqrt{2\pi}}\exp\left(\frac{\Delta^2}{4}\right)K_0\left(\frac{\Delta^2}{4}\right)\,. \tag{20b}$$

The functions $K_\nu(x)$ are modified Bessel functions. Since $\delta_{\text{PS}-AB}$ is in the HOM phase (20a) results from (11). For (20b) we exploited the explicit integrability of (18) at half filling, i.e. for $|\tilde{\mu}| \leq |\Delta|$. Explicit integration in (19) leads to the condition for the straight line

$$\frac{\Delta^2}{4}\exp\left(\frac{\Delta^2}{4}\right)K_1\left(\frac{\Delta^2}{4}\right) - \exp\left(-\frac{\tilde{\mu}_{\text{PS}-AB}^2}{2}\right) = \sqrt{2\pi}\left(\frac{U}{2}\delta_{\text{PS}-AB}^2 - \tilde{\mu}_{\text{PS}-AB}\delta_{\text{PS}-AB}\right)\,. \tag{21}$$

Numerical evaluation of (20) and (21) leads to the first order transition line (solid curve) in fig. 5. For comparison the second order transition line (dotted curve) $\chi_0^{-1}(-1)$ is plotted, too. This curve consists of the points ($\delta_{\text{div}-AB}, U$) which lie by construction to the left of the solid curve (cf. fig. 4). A point in the region PS-AB/HOM stands for a phase separated system consisting of a hole-rich homogeneous phase (volume fraction $v_1$) and a hole-free AB-CDW (volume fraction $1 - v_1$) such that the average doping equals $\delta$, i.e. $v_1 = \delta/\delta_{\text{PS}-AB}$

### 3.3 First order transition IP$\longleftrightarrow$AB-CDW

Although for most values of the interaction $U$ the first order transition line turns out to lie to the right of the second order transition line for optimised parameter of incommensurability there is an intervall $U \in [U_{\text{IPL}}, U_{\text{IPH}}] = [0.572, 1.914]$ where the situation is inverse (cf. fig. 6). Approaching half filling from large doping $\chi(\eta_{\text{opt}})$ diverges in this interval before the first order transition HOM$\longleftrightarrow$AB-CDW occurs. The existence of this interval was not yet found in ref. [43] due to an incorrect evaluation of the Maxwell condition.

The occurence of $U_{\text{IPL}}$ can be understood already in the limit $\delta \to 0$. Expanding (21) to order $\delta_{\text{PS}-AB}^2$ using $K_1(x) = 1/x + \mathcal{O}(x)$ yields $\Delta^2 = 4\pi\delta_{\text{PS}-AB}^2(1 + U/\sqrt{2\pi})$. From this one deduces with the help of (20)

$$\delta_{\text{PS}-AB} = \frac{1}{\sqrt{\pi}}\frac{\exp(-\sqrt{2\pi}/U - C/2 + \ln(2)/2)}{\sqrt{1 + U/\sqrt{2\pi}}}\,. \tag{22}$$

Comparing (22) with (14) and (13) for very small $U$ one realizes that the following sequence holds: $\delta_{\text{PS}-AB} > \delta_{\text{IP}} > \delta_{\text{AB}}$. This implies that phase separation dominates for $U \to 0$ at $T = 0$ [43]. Exactly the same observation was also made by van Dongen in the extended Hubbard model because the fluctuation induced renormalisation factor is equal in fact for all three relation (13), (14) and (22) [45].

The denominator in (22) leads to a change of the sequence at around $U \approx 0.56$: $\delta_{\text{IP}} > \delta_{\text{PS-AB}} > \delta_{\text{AB}}$. Thus one understands why in the exact calculation the incommensurate phase dominates the system above some particular value $U_{\text{IPL}} \approx 0.572$ of the interaction. Our calculations revealed that the sequence changes back to $\delta_{\text{PS-AB}} > \delta_{\text{IP}}$ at a higher value $U_{\text{IPH}} \approx 1.914$ of the interaction and stays like that for $U \to \infty$.

If the system is in the incommensurate phase for $U \in [U_{\text{IPL}}, U_{\text{IPH}}]$ at some finite doping and in the AB-CDW at half filling the question remains where the transition takes place and which kind of transition it is. To give an answer to this question one needs the ground state energy of the incommensurate phase, i.e. one has to know of which structure the order parameter is. Note that this is a question without a straightforward answer even though the interaction induces only Hartree terms [51]. In the following we will discuss a new ansatz for the spatial structure of the order parameter which goes beyond the assumption of charge density waves characterised by well-defined wave vectors. In contrast to the case of spin ordering in the Hubbard model it is not possible to resort to a spiral ordering with sufficiently high symmetry which can be resolved by standard methods.

For the spatial dependence of the order parameter $b(\vec{r}) := \langle \hat{n}_{\vec{r}} \rangle - n$ we use the product ansatz

$$b(\vec{r}) = b_0 \prod_{i=1}^{d} u_i(r^{(i)}) \tag{23}$$

where the functions $u_i$ take the values $\pm 1$ with equal frequency over the whole lattice and $r^{(i)}$ is the component $i$ of $\vec{r}$. So, the spatial dependence of $b(\vec{r})$ in each direction is independent from the dependence in the other directions. This form is most suitable for the description in the limit $d \to \infty$. Note that no periodicity of the $u_i(r^{(i)})$ is required.

In appendix A it is shown that the ground state energy depends on $b(\vec{r})$ only via the amplitude $b_0$ and the relative frequency $h$ on the whole lattice that $b(\vec{r})$ has the *same* sign on the *adjacent* sites. Unless $h$ is strictly unity the average of $b(\vec{r})$ over the whole lattice vanishes. Note that $h$ can take all values between 0 and 1 in the limit $d \to \infty$; in finite dimensions the values of $h$ are multiples of $1/(2d)$ on hypercubic lattices. The amplitude $b_0$ and the relative frequency $h$ are the only parameters which are relevant for the influence of a spatial structure of the form (23) on the energy.

In order to compute the energy functional for $b(\vec{r})$ most easily we exploit the freedom of choosing the functions $u_i$. Keeping $h$ constant we set $u_i(r^{(i)}) := 1$ for $i \leq hd$ and $u_i(r^{(i)}) := (-1)^{r^{(i)}}$ for $i > hd$. This choice is a particular realization of the relative frequency $h$ which restores periodicity and can be characterised by the wave vector

$$\mathbf{Q}_h = (\underbrace{0, 0, \ldots, 0}_{hd}, \underbrace{\pi, \pi, \ldots, \pi}_{(1-h)d})^{\dagger} . \tag{24}$$

With the help of (24) one can define a relation between $\eta$ and $h$, namely $\eta_{\mathbf{Q}_h} = 2h - 1$.

Since the ground state energy $E$ of the IP characterised by (23) is independent from how the relative frequency $h$ is realized one can compute $E$ for the CDW belonging to $\mathbf{Q}_h$. In this way one obtains the ground state energy for a large class of order parameters most of which do not display any periodicity from a simple calculation involving only $2 \times 2$ matrices. This calculation, which can be found in appendix A, yields with $\Delta = -\eta U b_0$

$$E(n, \Delta, \eta) = \frac{Un^2}{2} - \frac{\Delta^2}{2\eta U} + \int_{-\infty}^{\infty} dv \, \frac{\exp\left(-\frac{v^2}{1-\eta}\right)}{2\sqrt{\pi(1-\eta)}} \left[ \frac{\sqrt{v^2 + \Delta^2}}{\text{sgn}(v)} \, \text{erf}(P) - \frac{\sqrt{1+\eta}}{\sqrt{\pi}} \exp(-P^2) \right] \quad (25a)$$

$$P := \frac{\tilde{\mu} - \text{sgn}(v)\sqrt{v^2 + \Delta^2}}{\sqrt{1+\eta}} \qquad \delta = -\int_{-\infty}^{\infty} dv \, \frac{\exp\left(-\frac{v^2}{1-\eta}\right)}{2\sqrt{\pi(1-\eta)}} \, \text{erf}(P) \, . \quad (25b)$$

The actual energy is found from (25) by minimisation with respect to $\Delta$ and $\eta$. The generic result for $U = 1.2 \in [U_{\text{IPL}}, U_{\text{IPH}}]$ is depicted in the solid curve in fig. 4. Indeed, the solid curve lies always below the dashed-dotted curve which belongs to the AB-CDW ($\eta = -1$). But the minimisation with respect to $\eta$ does not remove the non-convexity. This means that there will still occur phase separation. For illustration the dopings $\delta_{\text{div-AB}}$ (divergence of $\chi(-1)$) and $\delta_{\text{PS-AB}}$ (start of AB phase separation) are depicted which are relevant when only $\eta = -1$ is considered. The doping $\delta_{\text{div-IP}}$ depicts where the susceptibility $\chi(\eta_{\text{opt}})$ diverges. A second order transition to the incommensurate phase occurs which is stable from $\delta_{\text{div-IP}}$ to $\delta_{\text{PS-IP}}$. At $\delta_{\text{PS-IP}}$ the first order transition from the IP to the AB-CDW sets in, i.e. one constructs the convex hull to the solid line. The convex hull to the dashed-dotted line is no longer relevant. The $\eta$ values which are relevant in the IP are very close to those shown in fig. 3 (b). So we do not display them in a separate figure.

Summarising the results of this section we conclude that for $U \in [U_{\text{IPL}}, U_{\text{IPH}}]$ the system is in the homogeneous phase at large doping; on decreasing doping it passes to the incommensurate phase by a second order transition and finally by a first order transition to the AB-CDW at half filling. The resulting $T = 0$ phase diagram is plotted in fig. 6. For $U < U_{\text{IPL}}$ or $U > U_{\text{IPH}}$ no IP occurs and the system passes by a first order transition directly from the HOM phase to the AB-CDW at $\delta = 0$.

## 4  Phase diagram at finite temperatures

In principle, the arguments of the preceding section can be used to derive a phase diagram at $T > 0$, too. Yet there arise some new aspects. The energy as function of the doping is smoothed by thermal fluctuations so that the regions of phase separation are shifted to higher values of the interaction and one finds also a second order transition HOM $\longleftrightarrow$ PS-AB/HOM. Furthermore the incommensurate phase becomes more important at finite temperatures [43].

## 4.1 Second order transitions

First we investigate the stability of the homogeneous phase, which is stable at low interaction and low filling, on increasing interaction and filling. Starting from (9c) one can perform one integration after the substitutions $\epsilon_2 = u\sqrt{2(1-\eta^2)} + \eta\epsilon_1$ or $\epsilon_1 = u\sqrt{2(1-\eta^2)} + \eta\epsilon_2$, respectively. With the following representation of the Dawson-function

$$D(\omega) = (4\pi)^{(-1/2)} \text{vp} \int_{-\infty}^{\infty} \exp(-u^2)/(\omega+u)\, du \qquad (26)$$

(vp for principal value) one obtains

$$\chi_0(\eta) = \int_{-\infty}^{\infty} \frac{d\omega \exp(-\omega^2/2)}{\sqrt{\pi(1-\eta^2)}} D(\omega\sqrt{(1-\eta)/(2(1+\eta))})\, (f_{\text{F}}(-v) - f_{\text{F}}(v))\ . \qquad (27)$$

As at $T = 0$ we compute the value of the parameter of incommensurability $\eta$ at which $\chi(\eta)$ in (10) diverges first. The results for $\chi(-1)$ and for $\chi(\eta_{\text{opt}})$ at $T = 0.1$ are presented in fig. 7(a). Comparison with the corresponding fig. 3(a) at zero temperature shows that at $T > 0$ the intersections are shifted to larger interaction on the divergence line of $\chi(\eta_{\text{opt}})$. The corresponding values of $\eta_{\text{opt}}$ at $T = 0.1$ are depicted in fig. 7(b). At $\delta_{\text{START}}$ the value of $\eta_{\text{opt}}$ goes continuously to $-1$ whereas at $\delta_{\text{END}}$ the value of $\eta_{\text{opt}}$ jumps discontinuously back to $-1$. Note that finite temperatures lower the values of $\eta_{\text{opt}}$; its maximum value is attained at $T = 0$.

The interval of incommensurability $[\delta_{\text{START}}, \delta_{\text{END}}]$ becomes smaller on increasing temperature. At $T_{\text{Imax}} \simeq 0.1826$ the interval ceases to exist with $\delta_{\text{START}} = \delta_{\text{END}}$. The values of $\delta_{\text{START}}$ and of $\delta_{\text{END}}$ can be found in fig. 8. The dashed and the dotted curve enclose the area in the $T$-$\delta$ plane where the divergence of $\chi(\eta)$ occurs first for $\eta > -1$. The dotted curve represents the dopings where $\eta_{\text{opt}}$ approaches $-1$ continuously and the dashed curve represents the dopings where $\eta_{\text{opt}}$ jumps to $-1$.

In contrast to the situation at $T = 0$ there exists at $T > 0$ a region of dopings around half filling for which the AB-CDW is stable. Thus a complete stability analysis must investigate *where* the AB-CDW becomes unstable. One possible instability can be calculated from a generalised susceptibility which is given by the $2 \times 2$ density-density response matrix $\boldsymbol{\chi}(\eta)$. One has to sum the geometric RPA series in fig. 2 taking into account the matrix form of the propagator in (17). This yields the extension of (10) to finite order parameter

$$\det(\boldsymbol{\chi}^{-1}(\eta,\Delta)) = \det\left(\boldsymbol{\chi}_0^{-1}(\eta) + U\eta\begin{bmatrix}1 & 0\\0 & -1\end{bmatrix}\right) = 0\ . \qquad (28)$$

The matrix in the $\eta$-term accounts for the fact that $\eta_{\vec{k}+\vec{Q}} = -\eta_{\vec{k}}$ which is relevant for the mode at $\vec{k} + \vec{Q}$. The matrix elements of

$$\boldsymbol{\chi}_0(\eta,\Delta) = \begin{bmatrix}\chi_0^{\text{d}}(\eta,\Delta) & \chi_0^{\text{od}}(\eta,\Delta)\\ \chi_0^{\text{od}}(\eta,\Delta) & \chi_0^{\text{d}}(-\eta,\Delta)\end{bmatrix} \qquad (29)$$

are derived in appendix B. The determinant appears in (28) in order to detect when $\boldsymbol{\chi}(\eta,\Delta)$ diverges, i.e. when one of the eigen values of $\boldsymbol{\chi}^{-1}(\eta,\Delta)$ becomes zero. The AB-CDW is stable if and only if both eigen values are positive. Starting at half filling and increasing $\delta$ we look for the value of $\eta$ at which $\chi(\eta, \Delta)$ diverges first. By this procedure the dotted curve in fig. 9 is obtained.

Coming from $\delta = 0$ in fig. 9 one realizes that the solid line **ab** precedes the dotted curve. This solid line marks the second order transition from the AB-CDW with $\eta = -1$ to the incommensurate CDW with $\eta > -1$ at finite amplitude of the CDWs. Note that the instability of the AB-CDW towards a deviation of $\eta$ from -1 is not equivalent to its instability towards the superposition with a CDW with infinitesimal amplitude and different value of $\eta$. The latter instability is signalled by the divergence of $\chi(\eta, \Delta)$. The former is determined from the minimisation of the free energy given in (33) in the next subsection in the limit $\eta \to -1$. Relying on our ansatz (23) we find that the instability of the AB-CDW towards a deviation of $\eta$ from -1 is the relevant instability since it precedes the divergence of $\chi(\eta, \Delta)$. This modifies our results presented previously [43].

## 4.2 First order transitions

In order to calculate first order transitions one has to construct the convex hull to a non-convex free energy as function of the filling. The free energy functional for the AB-CDW at $T > 0$ is given by

$$F(n, \Delta) = \frac{\Delta^2}{2U} - \frac{U}{2}n^2 + \mu n + T\int_{-\infty}^{\infty} d\omega \ln\left(1 - f_{\text{F}}(\text{sgn}(\omega)\sqrt{\omega^2 + \Delta^2})\right) \rho(\omega) , \qquad (30)$$

which can be constructed by integrating $\partial F(n, \Delta)/\partial n|_\Delta = \mu$ and $\partial F(n, \Delta)/\partial \Delta|_n = 0$ or by using the relation between the grand canonical potential given by $F(n, \Delta) - \mu n$ and the generating functional $\Phi$ in the Baym/Kadanoff formalism [47]

$$F(n, \Delta) - \mu n = -\left(\Phi + T\text{Tr}\left(\ln(-\mathbf{G}) - \mathbf{\Sigma G}\right)\right) . \qquad (31)$$

Since the cusp in the ground state energy at half filling (see fig. 4) is rounded at $T > 0$ the Maxwell construction requires to determine *two* dopings: $\delta_1 < \delta_2$. The larger one, $\delta_2$, refers to the hole-rich HOM phase; the smaller one, $\delta_1$, to the hole-poor AB-CDW. Between $\delta_1$ and $\delta_2$ the convex hull is given by a double tangent with the slope $\mu_{\text{AVE}}$. So the conditions

$$\int_{1/2-\delta_2}^{1/2-\delta_1} dn\, \mu(n) = F(1/2 - \delta_1, 0) - F(1/2 - \delta_2, \Delta) = (\delta_2 - \delta_1)\mu_{\text{AVE}} \qquad (32\text{a})$$

$$\frac{\partial F}{\partial n}(1/2 - \delta_1, 0) = \mu_{\text{AVE}} \qquad \frac{\partial F}{\partial n}(1/2 - \delta_2, \Delta) = \mu_{\text{AVE}} \qquad (32\text{b})$$

have to be fulfilled for phase separation between the HOM phase and the AB-CDW to occur. They determine $\Delta, \delta_1$ and $\delta_2$ for given interaction and temperature.

A generic result is shown in fig. 10 at $T = 0.2$. (At relatively high temperatures it is not necessary to account for the appearance of the incommensurate phase as we will see below.)

Below $U \approx 5.0$ no phase separation occurs. We observe a second order transition between the HOM phase and the AB-CDW. The point **a** in fig. 10 is the critical point where the second order transition line meets the two first order transition lines given by $U(\delta_1)$ and $U(\delta_2)$. A point in the phase diagram 10 at a certain doping $\delta$ in the region PS-AB/HOM represents a phase mixture where the HOM phase and the AB-CDW coexists in spatially distinct domains. The AB-CDW occupies a region of volume fraction $v_1$ and the HOM phase a region of volume fraction $1 - v_1$ so that $\delta = v_1 \delta_1 + (1 - v_1)\delta_2$ holds.

It is instructive to keep track of the loci of the critical points in the $T$-$\delta$ plane. They are given from (32) in the limit $\delta_2 - \delta_1 \to 0$. Since they lie all on the second order transition line HOM$\longleftrightarrow$AB-CDW their interaction co-ordinate is given by $U = \chi^{-1}(\eta = -1)$. The solid curve in fig. 8 depicts the critical points in the $T$-$\delta$ plane.

From the origin to point **a** in fig. 8 the critical points of the AB phase separation lie *within* the intervals where the incommensurate phase is important. Let us define the tripel $(\delta_{\text{IP/PS}}, T_{\text{IP/PS}}, U_{\text{IP/PS}}) = (0.2668, 0.1825, 4.212)$ by the co-ordinates of this point **a**. Then we can make the following statements:

- Beyond point **a** phase separation dominates. For $T > T_{\text{IP/PS}}$ (or for $U > U_{\text{IP/PS}}$) the HOM$\longleftrightarrow$-AB-CDW phase separation prevails, i.e. essentially *no* incommensurate phases occur. In this sense the phase diagram in fig. 10 is generic.

- For $T < T_{\text{IP/PS}}$ (or for $U < U_{\text{IP/PS}}$) the incommensurate phase occurs. It will *mask* the critical point of the AB phase separation. The generic situation is shown in fig. 9 for $T = 0.1 < T_{\text{IP/PS}}$.

In order to investigate phase separation including the incommensurate phase given by the ansatz (23) the free energy functional

$$F(n, \Delta; \eta) = -\frac{\Delta^2}{2\eta U} - \frac{U}{2}n^2 + \mu n + T \int_{-\infty}^{\infty} d\omega \, \ln(1 - f_{\text{F}}(\omega)) \, \rho_\eta(\omega) \quad \text{with} \tag{33a}$$

$$\rho_\eta(\omega) = \int_{-\infty}^{\infty} \frac{du\,dv}{2\pi\sqrt{h(1-h)}} \exp\left(-\left[u^2/h + v^2/(1-h)\right]/2\right) \delta(\omega - \epsilon(u,v)) \tag{33b}$$

for general $\eta$ must be known. It is derived in appendix A. The minimisation of the free energy in (33) with respect to $\Delta$ and to $\eta$ determines these values. In analogy to the case $T = 0$ we construct the convex hull to $F(n, \Delta; \eta)$ by the Maxwell construction. Like at $T = 0$ we find a first order transition from the AB-CDW at low doping to the IP phase at higher doping. A region of phase separation PS-AB/IP occurs which is shown in fig. 9 between the curves **ad** and **bc**. Point **b** is a new critical point where the second order line **ab** between IP and AB-CDW splits into the two first order lines.

Comparing the case $T > 0$ in fig. 9 with the case at $T = 0$ in fig. 6 one finds that the region in which the pure IP phase is stable is considerably enlarged by finite temperatures.

This will become even more obvious by examination of the phase diagram at fixed interaction $U$. Up to now we discussed the phase diagram in the $U - \delta$ plane at fixed temperature which was the best way to understand the physical principles which govern the interplay of the different phases. On the other hand, it is experimentally very interesting to investigate the system at given interaction because the temperature is more easily tuned than the value of the interaction. The theoretical interest in the phase diagram at given interaction is kindled by the appearance of a re-entrant behaviour (see below).

We do not want to present here the whole line of argument for the derivation of the phase diagrams in the $T - \delta$ plane which is analogous to the one for the $U - \delta$ plane. We only state that for the $T - \delta$ plane the two limits in which the character of the phase is known are the limit $T \to \infty$ and the limit $T \to 0; \delta \to 0$. In the former limit the phase is homogeneous; in the latter limit the phase is the AB-CDW. They serve as the starting points of the stability analysis.

The results are shown in figs. 11(a)–(d). Depending on the value $U$ of the interaction there are four generic cases: $U \in [U_{\text{IP/PS}}, \infty]$, $U \in [U_{\text{IPH}}, U_{\text{IP/PS}}]$, $U \in [U_{\text{IPL}}, U_{\text{IPH}}]$ and $U \in [0, U_{\text{IPL}}]$.

One recognizes the meaning of $U_{\text{IP/PS}}$ which was defined analogously to $T_{\text{IP/PS}}$ with the help of fig. 8. For values of the interaction $U > U_{\text{IP/PS}}$ as in figure 11(a) the incommensurability is totally absent. For $U < U_{\text{IP/PS}}$ an IP appears.

Depending on the value of $U$ one encounters different scenarios for $T \to 0$. From fig. 6 one sees that for $U > U_{\text{IPH}}$ the system does not display the IP at $T = 0$. So it appears only at finite temperatures as depicted in fig. 11(b). For $U_{\text{IPH}} > U > U_{\text{IPL}}$, however, fig. 6 tells us that the IP appears together with the PS-AB/IP down to $T = 0$ as shown in fig. 11(c). For $U_{\text{IPL}} > U > 0$ in fig. 11(d) there is again a phase separated region (PS-AB/HOM) at $T = 0$ similar to the case for $U_{\text{IP/PS}} > U > U_{\text{IPH}}$ (fig. 11(b)).

Inspection of figs. 11(a)–(c) shows that re-entrant behaviour occurs since the phase transition lines are not always functions of $\delta$. (In fig. 11(c) the effect is hardly discernible in the plot.) For instance, in fig. 11(a) at $\delta \lesssim 0.3$ the system is in the HOM phase for temperatures $0 < T \lesssim 0.25$ and $T \gtrsim 0.5$ whereas for $0.25 \lesssim T \lesssim 0.5$ it forms a AB-CDW. The phase sequences which occur on increasing temperature can also be very sophisticated. In fig. 11(a) one sees that the system at $\delta \approx 0.29$ undergoes the transitions HOM$\longrightarrow$PS-AB/HOM$\longrightarrow$AB-CDW$\longrightarrow$HOM on raising $T$ from zero to 1. Another example is found in fig. 11(b) at $\delta \approx 0.165$ where the sequence of transitions on raising $T$ from zero to 0.6 is PS-AB/HOM$\longrightarrow$PS-AB/IP$\longrightarrow$IP$\longrightarrow$AB-CDW$\longrightarrow$IP$\longrightarrow$HOM.

Summarising the results of the section on finite temperatures we conclude that second order transitions become more important at $T > 0$. Phase separation is shifted to higher values of the interaction. The AB-CDW is stabilised for finite doping and exists not only at $\delta = 0$. Equally the relevance of the incommensurate phase is enhanced. For $U \in [U_{\text{IPH}}, U_{\text{IP/PS}}] = [1.914, 4.212]$ and for $U \in [0, U_{\text{IPL}}] = [0, 0.572]$ the incommensurate phase occurs at finite temperatures but *not* at zero temperature. Additionally we found re-entrant behaviour as function of the temperature in the sequence of phases at given doping and interaction.

## 5 Discussion

In this paper we discussed the phase diagram of the model of spinless fermions with repulsive nearest neighbour interaction on a hypercubic lattice in infinite dimensions. In the limit $d \to \infty$ the perturbation series can be resummed and one finds that the Hartree approximation becomes exact. This allows investigations in great detail of the phases and their spatial structure. In the whole parameter space of interaction, doping and temperature we examined the symmetry unbroken homogeneous phase (HOM), the AB charge density wave (AB-CDW) and the incommensurate phase (IP). We performed a complete stability analysis for the HOM and the AB-CDW relying on the RPA which is exact in the limit $d \to \infty$. For the spatial structure of the incommensurate phase we proposed a very general ansatz (23) in product form. We were able to show that within this framework the free energy of the IP depends only on the amplitude $b_0$ of the order parameter and on the parameter of incommensurability $\eta$ which is directly linked by $\eta = 2h - 1$ to the relative frequency $h$ that the order parameter has the same sign on two randomly chosen adjacent sites (see appendix A).

At zero temperature we found that the system is in the HOM phase for large dopings and enters the AB-CDW on decreasing doping. For small values and for large values of the interaction this transition is of first order and implies a region of phase separation between the HOM phase and the AB-CDW (PS-AB/HOM). In the intermediate range $U \in [U_{\text{IPL}}, U_{\text{IPH}}] = [0.572, 1.914]$ the sequence of transitions for $\delta \to 0$ is more sophisticated since the system enters first the IP via a second order transition (see fig. 6). Then the system undergoes a first order transition IP$\longleftrightarrow$AB-CDW with phase separation between the IP phase and the AB-CDW (PS-AB/IP).

Finite temperatures favour second order transitions; the first order transitions and the regions of phase separation are shifted to larger values of the interaction. For $T > 0$ the AB-CDW is stable for finite dopings, too. A very important result is that the region of the appearance of the IP is enlarged by finite temperatures. Besides the two important interaction values $U_{\text{IPL}}$ and $U_{\text{IPH}}$ a third one, $U_{\text{IP/PS}} = 4.212$, must be taken into account. For values of the interaction below $U_{\text{IP/PS}}$ the IP appears in a certain temperature range. If $U$ is not in the interval $[U_{\text{IPL}}, U_{\text{IPH}}]$ this temperature range does not extend down to $T = 0$. The three important interaction values $U_{\text{IPL}} < U_{\text{IPH}} < U_{\text{IP/PS}}$ divide the positive $U$-scale into four intervals of different generic behaviour. The generic phase diagrams are given in fig. 11.

Our results can also be used to gain a better understanding of the phase diagram of other models such as the Hubbard model and the extended Hubbard model. The argument why this transfer is reasonable comprises two points. The first is the similarity of the weak coupling expansion (Hartree approximation) of the (extended) Hubbard model to the treatment of spinless fermions presented here. Of course, there are even more phases present in the Hubbard model since there are charge and spin degrees of freedom. But the weak coupling calculation for charge density waves *or* spin density waves of a given wave vector are identical to the calculations for the corresponding charge density wave in the spinless fermion model except for a factor of 2 due to the presence of two fermion species in the Hubbard model. The second point concerns the influence of quantum fluctuations. They are suppressed in the spinless fermion model as treated here by the limit $d \to \infty$ but they are present in the Hubbard model. Fortunately van Dongen showed that the effect of these fluctuations can be accounted for in the weak coupling regime by a renormalisation $q$ of the order of 1 [52, 53]. Moreover, he showed that in the vicinity of half filling this renormalisation is the *same* for the different phases [45]. From these considerations we conclude that the Hubbard model with pure on-site interaction and the Hubbard model with pure nearest-neighbour interaction display in the weak coupling limit the same phase diagrams as we calculated except for a rescaling of the doping and of the interaction. The general extended Hubbard model should have a phase diagram which is in many aspects similar to the ones shown in the present work.

For instance, van Dongen's result $0.56 = U_{\text{on-site}} + 2U_{\text{nearest-neighbour}}$ in our units, i.e. the interaction values above which the IP is more favourable than the PS-AB/HOM, is in perfect agreement with $U_{\text{IPL}} = 0.572$ up to the factor 2 due to the spin degeneracy. The difference between 0.572 and 0.56 results from the fact that van Dongen compared the analogues of the asymptotic expressions (13) and (22) whereas we define $U_{\text{IPL}}$ from the numeric results in fig. 6.

At small but finite interactions the renormalisation due to quantum fluctuations is not only a simple constant. Nevertheless the qualitative picture remains the same as can be seen in a comparison to QMC results by Freericks and Jarrell [50]. They define the temperature ratio $T_I/T_N$ of the temperature $T_I$ at finite doping where the IP sets in and the Néel temperature $T_N$ at half filling where the AB phase (here: spin density wave) sets in. This ratio is found to be fairly constant at about 0.57. It is in good agreement to our results for the ratio of the temperature co-ordinate of the points **a** in figs. 11(b)–(d) to the temperature where the AB-CDW appears at half filling. We find 0.31 at $U = 2.5$, 0.51 at $U = 1$ and 0.56 at $U = 0.45$ (from figs. 11(b)-(d)). Naturally the agreement becomes excellent for smaller values of the interaction. In the comparison of interaction values a renormalisation due to fluctuations of the order of 3 should be kept in mind [50].

In contrast to the results of van Dongen, Freericks and Jarrell find incommensurate order and no sign of phase separation (although they cannot rule out the latter completely). These seemingly differing findings can be reconciled by our results. In the quantum Monte Carlo calculations the phases at $T = 0$ are found by extrapolation from finite temperature results, i.e. one investigates which phase occurs first on lowering the temperature from the paramagnetic phase (corresponding to the HOM phase). Looking at figs. 11(b)–(d) one notes that it is either the AB phase or the incommensurate phase that is found. There is (almost) no chance to find a sign of phase separation, especially since the divergence of the compressibility occurs behind the right phase separation line. Adopting, however, van Dongen's view point at $T = 0$ and $U \to 0$ fig. 6 clearly shows that AB phase separation (PS-AB/HOM) dominates. This resolves the seeming contradiction and adds to the understanding of the phase diagram in the more complicated case of the Hubbard model. The QMC result confirms our statement that finite temperatures favour the appearance of incommensurate phases also in the Hubbard model. We are convinced that the infinite dimensional Hubbard model displays also phase separation between the antiferromagnetic phase and an incommensurate spiral phase corresponding to PS-AB/IP.

Further investigations should incorporate the effect of fluctuations which implies the consideration of $1/d$ corrections for the spinless fermion model. This was done so far only at half filling [48] due to the high complexity of the phase diagram at finite doping.

We thank D. Vollhardt and P. van Dongen for helpful discussions. We are indepted to V. Janiš for valuable advice. Furthermore we gratefully acknowledge the financial support of the Deutsche Forschungsgemeinschaft in the Sonderforschungsbereich 341 (GSU) and the Stichting Fundamenteel Onderzoek der Materie (RV).

## Appendix A

In order to derive that the ground state energy per site, more generally the free energy $F$ per site, depends only on $b_0$ and $h$ if the order parameter $b(\vec{r})$ is assumed to be given by the ansatz (23) we consider the expansion of $F$ in powers of $b(\vec{r})$. The direct interaction part (representing $T\mathrm{Tr}\Sigma\mathbf{G} - \Phi$; cf. first two terms in (30)) is given by

$$F_{\mathrm{inter}} = (U/2)[(1-h)(n+b_0)(n-b_0) + (h/2)\left((n+b_0)^2 + (n-b_0)^2\right)] \tag{34a}$$

$$= (U/2)\left(n^2 - (1-2h)b_0^2\right) = (U/2)\left(n^2 + \eta b_0^2\right). \tag{34b}$$

The kinetic part (representing $-\mathrm{Tr}\ln(-\mathbf{G})$; cf. last term in (30)) results from an expansion in the inhomogeneous chemical potential given by $Ub(\vec{r})$

$$F_{\mathrm{kin}} - F_0 = T\sum_{\omega_n}\sum_{n=1}^{\infty} \frac{U^n}{n} S_n \quad \text{with}$$

$$S_n := \frac{1}{N}\sum_{\vec{r}_1,\vec{r}_2,\ldots,\vec{r}_n} b(\vec{r}_1)g(\vec{r}_1-\vec{r}_2)b(\vec{r}_2)g(\vec{r}_2-\vec{r}_3)\ldots b(\vec{r}_n)g(\vec{r}_n-\vec{r}_1) \tag{35}$$

where $g(\vec{r})$ is the propagator in real space. Its frequency dependence is omitted since all propagators are used at a fixed frequency $\omega_n$ due to the static character of the Hartree theory. Based on the following representation of the propagator $g(\omega_n, \vec{k})$ in momentum space

$$g(\omega_n, \vec{k}) = (i\omega_n - \epsilon_{\vec{k}} + \tilde{\mu})^{-1} = -i\int_0^{\infty} d\lambda \exp\left(i(i\omega_n - \epsilon_{\vec{k}} + \tilde{\mu})\lambda\right) \tag{36}$$

for $\omega_n > 0$ (for $\omega_n < 0$ one uses $i\lambda \to -i\lambda$) the propagator $g(\vec{r})$ in real space can be written as

$$g(\vec{r}) = \int \frac{d^d k}{(2\pi)^d} g(\omega_n, \vec{k}) \exp\left(i\vec{k}\vec{r}\right) = -i \int_0^\infty d\lambda \, \exp\left(-\lambda \omega_n + i\tilde{\mu}\lambda\right) \prod_{j=1}^d f(r^{(j)}, \lambda) \qquad (37a)$$

$$f(r, \lambda) := \int_{-\pi}^{\pi} \frac{dk}{2\pi} \exp\left(i\lambda\sqrt{2/d}\cos(k) + ikr\right) \qquad (37b)$$

$$= \delta(r) + i(\lambda/2)\sqrt{2/d}(\delta(r+1) + \delta(r-1)) + \mathcal{O}(d^{-1}) \, . \qquad (37c)$$

The superscript in $r^{(j)}$ stands for the $j$-component of the vector $\vec{r}$. The representation (37) together with (23) allows to write the essential part of $S_n$ as a product

$$S_n = b_0^n \int_0^\infty d\lambda_1 \ldots d\lambda_n e^{-(\omega_n - i\tilde{\mu})(\lambda_1 + \ldots + \lambda_n)} (-i)^n P^{(n)}(\{\lambda_i\}) \qquad (38a)$$

$$P^{(n)}(\{\lambda_i\}) := \prod_{j=1}^d P_j^{(n)}(\{\lambda_i\}) \quad \text{with} \qquad (38b)$$

$$P_j^{(n)}(\{\lambda_i\}) := \frac{1}{L} \sum_{r_1, r_2, \ldots, r_n} u_j(r_1) f(r_1 - r_2, \lambda_1) u_j(r_2) f(r_2 - r_3, \lambda_2) \ldots u_j(r_n) f(r_n - r_1, \lambda_n), \qquad (38c)$$

where the system size $N = L^d$ is split up assuming a hypercubic shape of the system. The expression in (38) is still exact in any dimension. Inserting the expansion (37c) one finds for odd $n$ that already the leading order $d^0$ vanishes since $b(\vec{r})$ vanishes on averaging over the lattice $(1/N) \sum_{\vec{r}} b(\vec{r}) = 0$. For even $n$ the leading order of (37c) yields 1. Now we apply an argument for $d \to \infty$ similar to the one used in the derivation of the central limit theorem expanding the logarithm of $P^{(n)}(\{\lambda_i\})$

$$\ln\left(P^{(n)}(\{\lambda_i\})\right) = \sum_{j=1}^d \ln\left(1 - \frac{1}{2dL} \sum_{r_1; a=\pm 1} \sum_{n \geq l' > l \geq 1} \lambda_l \lambda_{l'} u_j^{n-l'+l}(r_1) u_j^{l'-l}(r_1 + a) + \mathcal{O}\left(d^{-2}\right)\right) \qquad (39a)$$

$$= -\sum_{n \geq l' > l \geq 1} \lambda_l \lambda_{l'} \left(h + (1-h)(-1)^{(l'-l)}\right) + \mathcal{O}(d^{-1}) \, . \qquad (39b)$$

In the last step we used $u_j^n = 1$ for even $n$ and $u_j^{l-l'}(r_1) u_j^{l'-l}(r_1 + a) = 1$ for even $l' - l$. For odd $l' - l$ the expression $u_j^{l-l'}(r_1) u_j^{l'-l}(r_1 + a)$ takes the value 1 with the relative frequency $h$ and the value $-1$ with the relative frequency $1 - h$.

The crucial point in (39b) is that the result depends for $d \to \infty$ only on $h$ and not on further details of the functions $u_j$. Via (38) the free energy functional depends only on $b_0$ and on $h$ (or equivalently on $\eta = 2h - 1$).

We turn now to the explicit derivation of (19). Starting from

$$\boldsymbol{G}(i\omega_n, \vec{k}; h) = \begin{bmatrix} i\omega_n - \epsilon_{\vec{k}} + \tilde{\mu} & -\Delta \\ -\Delta & i\omega_n - \epsilon_{\vec{k}+\vec{Q}_h} + \tilde{\mu} \end{bmatrix}^{-1} \qquad (40)$$

with $\Delta := (1 - 2h)b_0 U = -\eta b_0 U$, it is appropriate to split $\epsilon_{\vec{k}} = u_{\vec{k}} + v_{\vec{k}}$ with

$$u_{\vec{k}} := -\sqrt{\frac{2}{d}} \sum_{n=1}^{hd} \cos(k_n) \qquad v_{\vec{k}} := -\sqrt{\frac{2}{d}} \sum_{n=hd+1}^{d} \cos(k_n) \qquad (41)$$

so that $\epsilon_{\vec{k}+\vec{Q}_h} = u_{\vec{k}} - v_{\vec{k}}$ holds. The one-particle energy resulting from (40) is $\epsilon(u,v) := u + \text{sgn}(v)\sqrt{v^2 + \Delta^2}$. The distribution of $u$ and $v$ is again easily found via the central limit theorem which permits to compute the DOS $\rho_\eta(\omega)$

$$\rho_\eta(\omega) = \int_{-\infty}^{\infty} \frac{du\, dv}{2\pi\sqrt{h(1-h)}} \exp\left(-\left[u^2/h + v^2/(1-h)\right]/2\right) \delta(\omega - \epsilon(u,v)) . \qquad (42)$$

Performing the $u$-integration and substituting $h$ by $(\eta+1)/2$ yields an expression for $\rho_\eta(\omega)$ which leads together with (34b) to the free energy

$$F(n, \Delta; \eta) = -\frac{\Delta^2}{2\eta U} - \frac{U}{2}n^2 + \mu n + T\int_{-\infty}^{\infty} d\omega \ln\left(1 - f_{\text{F}}(\omega)\right) \rho_\eta(\omega) . \qquad (43)$$

Note that $f_{\text{F}}(\omega) = 1/[1 + \exp(\beta(\omega - \tilde{\mu}))]$ In the limit $T \to 0$ one obtains the ground state energy in (19).

# Appendix B

The matrix elements in (29) are calculated from a single dressed bubble using (17)

$$\chi_0^{\text{d}}(\eta) = -\frac{T}{N} \sum_{\omega_n; \vec{k}'} \frac{\left[i\omega_n + \epsilon_{\vec{k}+\vec{k}'}\right]\left[i\omega_n + \epsilon_{\vec{k}'}\right] + \Delta^2}{\left[(i\omega_n)^2 - \epsilon_{\vec{k}'+\vec{k}}^2 - \Delta^2\right]\left[(i\omega_n)^2 - \epsilon_{\vec{k}'}^2 - \Delta^2\right]}$$

$$\chi_0^{\text{od}}(\eta) = -\frac{T}{N} \sum_{\omega_n; \vec{k}'} \frac{\Delta\left[i\omega_n + \epsilon_{\vec{k}+\vec{k}'}\right] + \Delta\left[i\omega_n + \epsilon_{\vec{k}'}\right]}{\left[(i\omega_n)^2 - \epsilon_{\vec{k}'+\vec{k}}^2 - \Delta^2\right]\left[(i\omega_n)^2 - \epsilon_{\vec{k}'}^2 - \Delta^2\right]} , \qquad (44)$$

where we write $i\omega_n$ for $i\omega_n + \tilde{\mu}$ for the sake of shortness. The sums $\sum_{\vec{k}}$ run over half the Brillouin zone ($\epsilon_{\vec{k}} < 0$) in order to avoid double counting. Performing the sum of Matsubara-frequencies $\omega_n$ and using (4) for the $\vec{k}$-sum yields

$$\chi_0^{\text{d}}(\eta) = \text{vp} \int_{-\infty}^{\infty} d\epsilon_1 d\epsilon_2 \rho(\epsilon_1, \epsilon_2; \eta) \frac{\epsilon_1^2 + \epsilon_1\epsilon_2 + 2\Delta^2}{z'(\epsilon_2^2 - \epsilon_1^2)} \left(f_{\text{F}}(-z') - f_{\text{F}}(z')\right) \qquad (45a)$$

$$\chi_0^{\text{od}}(\eta) = \text{vp} \int_{-\infty}^{\infty} d\epsilon_1 d\epsilon_2 \rho(\epsilon_1, \epsilon_2; \eta) \frac{\Delta\left(f_{\text{F}}(-z') + f_{\text{F}}(z')\right)}{\epsilon_2^2 - \epsilon_1^2} , \qquad (45b)$$

where vp stands for the principal value and $z' := \sqrt{\epsilon_1^2 + \Delta^2}$. The partial fraction expansion

$$\text{vp}\left(\frac{2}{\epsilon_1^2 - \epsilon_2^2}\right) = \text{vp}\left(\frac{1}{\epsilon_1}\right)\text{vp}\left(\frac{1}{\epsilon_1 - \epsilon_2} + \frac{1}{\epsilon_1 + \epsilon_2}\right) - \pi^2 \delta(\epsilon_1)\left(\delta(\epsilon_1 - \epsilon_2) + \delta(\epsilon_1 + \epsilon_2)\right) \qquad (46)$$

is useful to simplify the integration in (45). After the substitution $\epsilon_2 = u\sqrt{2(1-\eta^2)} + \eta\epsilon_1$ the $u$-integration is mapped on the representation (26) of the Dawson-function and one obtains finally

$$\chi_0^{\text{d}}(\eta) = \int_{-\infty}^{\infty} \frac{\exp(-\omega^2/2)}{\omega z \sqrt{\pi(1-\eta^2)}} \left[z^2\, \text{D}(\omega_-) + \Delta^2\, \text{D}(\omega_+)\right] F_-(z) d\omega - \frac{\pi\Delta}{2\sqrt{1-\eta^2}} F_-(\Delta) \qquad (47a)$$

$$\chi_0^{\mathrm{od}}(\eta) \;=\; \Delta \int_{-\infty}^{\infty} \frac{\exp(-\omega^2/2)}{\omega\sqrt{\pi(1-\eta^2)}} \left[\mathrm{D}(\omega_-) + \mathrm{D}(\omega_+)\right] F_+(z)d\omega - \frac{\pi\Delta}{2\sqrt{1-\eta^2}} F_+(\Delta) \quad \text{with} \quad (47\mathrm{b})$$

$$\omega_\pm \;:=\; \omega\sqrt{(1\pm\eta)/(2(1\mp\eta))} \qquad F_\pm(z) \;:=\; f_{\mathrm{F}}(-z) \pm f_{\mathrm{F}}(z) \qquad z \;:=\; \sqrt{\omega^2 + \Delta^2}\,. \quad (47\mathrm{c})$$

# Figure captions

Figure 1: First three contributions to the Baym/Kadanoff generating functional $\Phi$: (a) leading order $d^0$; (b) and (c) order $d^{-1}$.

Figure 2: Diagrammatic representation of (a) the Dyson equation (b) the general Bethe-Salpether equation and (c) the RPA series for the density-density response function $\chi(\vec{k})$ resulting from (b).

Figure 3: (a) solid curve: interaction values where $\chi(\eta_{\text{opt}})$ diverges at $T=0$; dotted curve: interaction values where $\chi(-1)$ diverges. (b) the corresponding values of $\eta_{\text{opt}}$, for $0 < \delta < \delta_{\text{END}} = 0.308$ holds $\eta_{\text{opt}} > -1$.

Figure 4: Ground state energies at $U = 1.2$, for the sake of clarity a linear function of $\delta$ is added ($a \approx 0.5561, b \approx 0.1946$); dotted curve : homogeneous phase (HOM); dashed-dotted curve: AB charge density wave (AB-CDW); solid curve: incommensurate phase with optimised $\eta = \eta_{\text{opt}}$ (IP). Divergence of $\chi(-1)$ at $\delta_{\text{div-AB}}$; divergence of $\chi(\eta_{\text{opt}})$ at $\delta_{\text{div-IP}}$. The Maxwell construction for the phase separation HOM$\longleftrightarrow$AB-CDW is a straight line between the points at $\delta = \delta_{\text{PS-AB}}$ and $\delta = 0$, the Maxwell construction for the phase separation IP$\longleftrightarrow$AB-CDW is a straight line between the points at $\delta = \delta_{\text{PS-IP}}$ and $\delta = 0$. The pure IP is present between $\delta_{\text{PS-IP}}$ and $\delta_{\text{div-IP}}$.

Figure 5: Phase separation between the HOM phase and the AB-CDW at $T = 0$. The phase separated region is marked PS-AB/HOM. Dotted curve: interaction values where $\chi(-1)$ diverges; solid curve: PS-AB/HOM boundary. The pure AB-CDW is only stable at half-filling ($\delta = 0$).

Figure 6: Phase diagram at $T = 0$: HOM: homogeneous phase, AB-CDW: AB charge density wave, IP: incommensurate phase, PS-AB/HOM: phase separation between HOM and AB-CDW, PS-AB/IP: phase separation between IP and AB-CDW. Solid curves: first order transition HOM⟷AB-CDW; dashed curve: first order transition IP⟷AB-CDW; dashed-dotted curve: second order transition HOM⟷IP.

Figure 7: (a) solid curve: interaction values where $\chi(\eta_{\text{opt}})$ diverges at $T = 0.1$; dotted curve: interaction values where $\chi(-1)$ diverges. (b) the corresponding values of $\eta_{\text{opt}}$, for $\delta \in [\delta_{\text{START}}, \delta_{\text{END}}] = [0.055, 0.319]$ holds $\eta_{\text{opt}} > -1$.

Figure 8: Relevance of the incommensurate phase: the dashed and the dotted curve enclose the region where $\eta_{\text{opt}} > -1$ is possible. The dotted curve shows the points of the continuous transition $\eta_{\text{opt}} \to -1$ on leaving the enclosed region whereas the dashed curve depicts the points where the transition $\eta_{\text{opt}} \to -1$ is discontinuous. The solid curve depicts the $\delta$ and $T$ co-ordinates of the critical points of the AB phase separation. The co-ordinates of the intersection **a** define $\delta_{\text{IP/PS}}$ and $T_{\text{IP/PS}}$.

Figure 9: Generic phase diagram (solid lines) at finite temperature $T = 0.1 < T_{\text{IP/PS}}$. Besides the nonsegregated phases AB-CDW, IP and HOM regions of phase separation occur between the AB-CDW and the incommensurate phase (PS-AB/IP) and between the AB-CDW and the homogeneous phase (PS-AB/HOM). The point **b** is tricritical. The PS-AB/IP starts there: curves above point **b** mark first order transitions. The curves **ab** and **ac** mark second order transitions.
The dashed and the dotted line show the phase diagram if the new ansatz for the IP were not taken into account. The dotted curve marks a second order transition from AB-CDW to IP calculated from (28). Note the significant difference between this curve and the curve **ab**. The dashed curve marks the beginning of a first order transition from AB-CDW to HOM.

Figure 10: Generic phase diagram at finite temperature $T = 0.2 > T_{\text{IP/PS}}$. Solid curve below the critical point **a**: second order transition HOM⟷AB-CDW; solid curves above point **a**: first order transition HOM⟷AB-CDW, left branch $U(\delta_1)$, right branch $U(\delta_2)$ (for $\delta_1, \delta_2$ see explanation to eq. (32)).

Figure 11: Generic phase diagrams at fixed interaction. (a) $U = 5.0 > U_{\text{IP/PS}}$: point **a** is the critical point where the second order transition line (above **a**) and the first order transition lines (below **b**) meet.
(b) $U = 2.5 \in [U_{\text{IPH}}, U_{\text{IP/PS}}]$; (c) $U = 1.0 \in [U_{\text{IPL}}, U_{\text{IPH}}]$; (d) $U = 2.5 \in [0, U_{\text{IPL}}]$. The points and the lines between them in figs. (b)-(d) have the same meaning as in fig. 9.

(a) 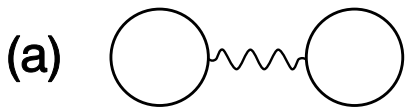

(b) 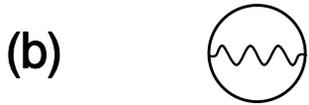

(c) 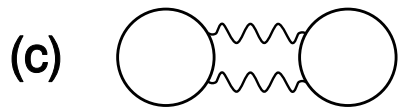

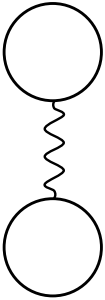 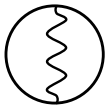 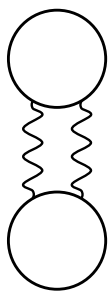

(a)　(b)　(c)

(a) 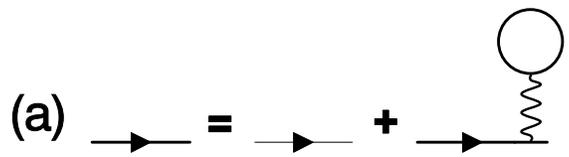

(b) 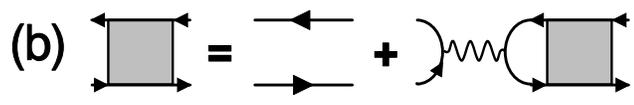

(c) 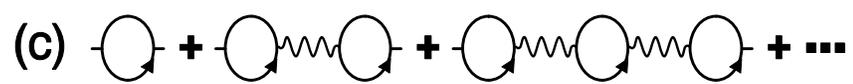

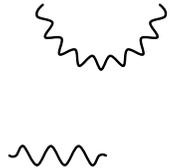 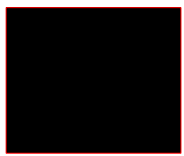

(a)

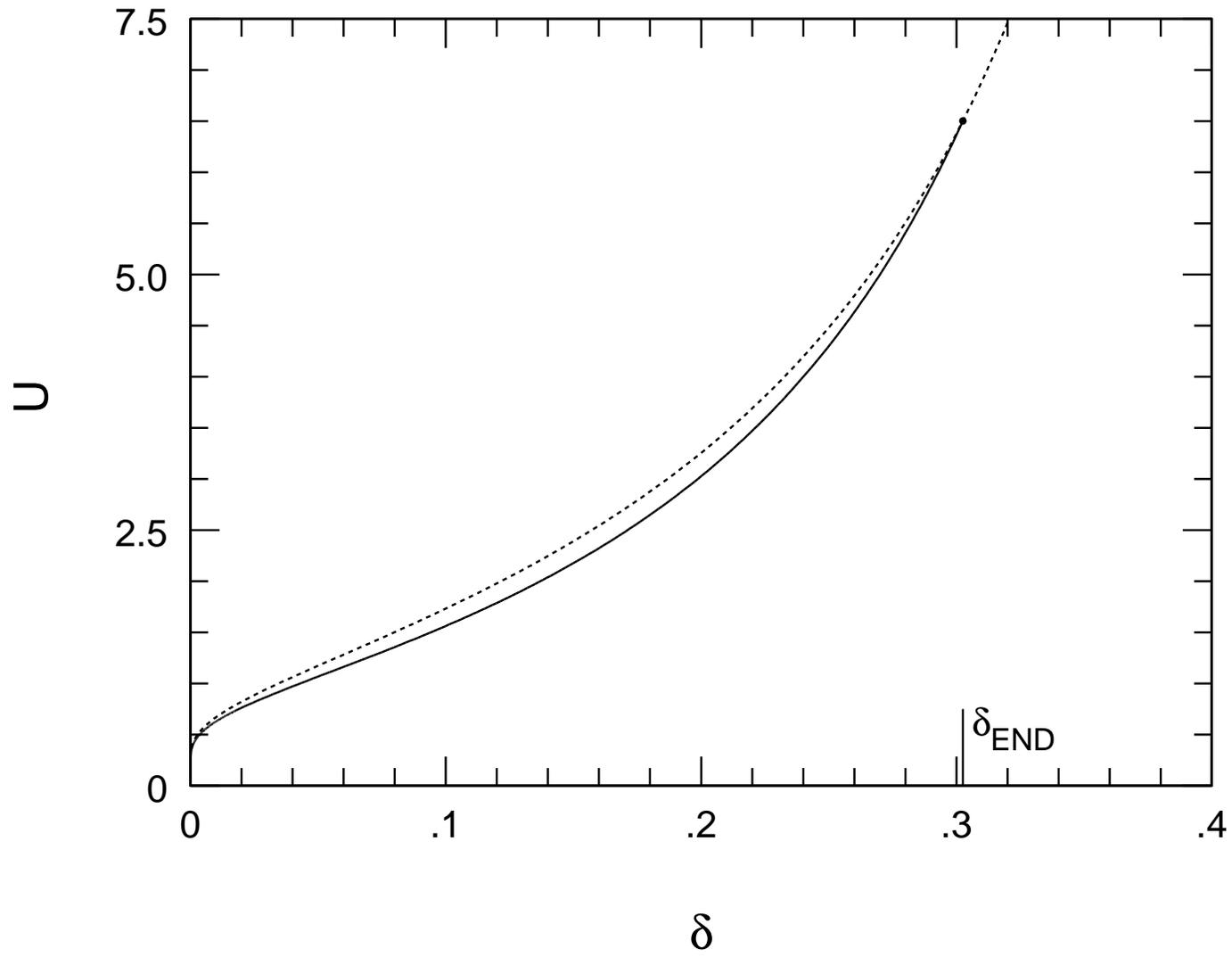

$\delta_{END}$

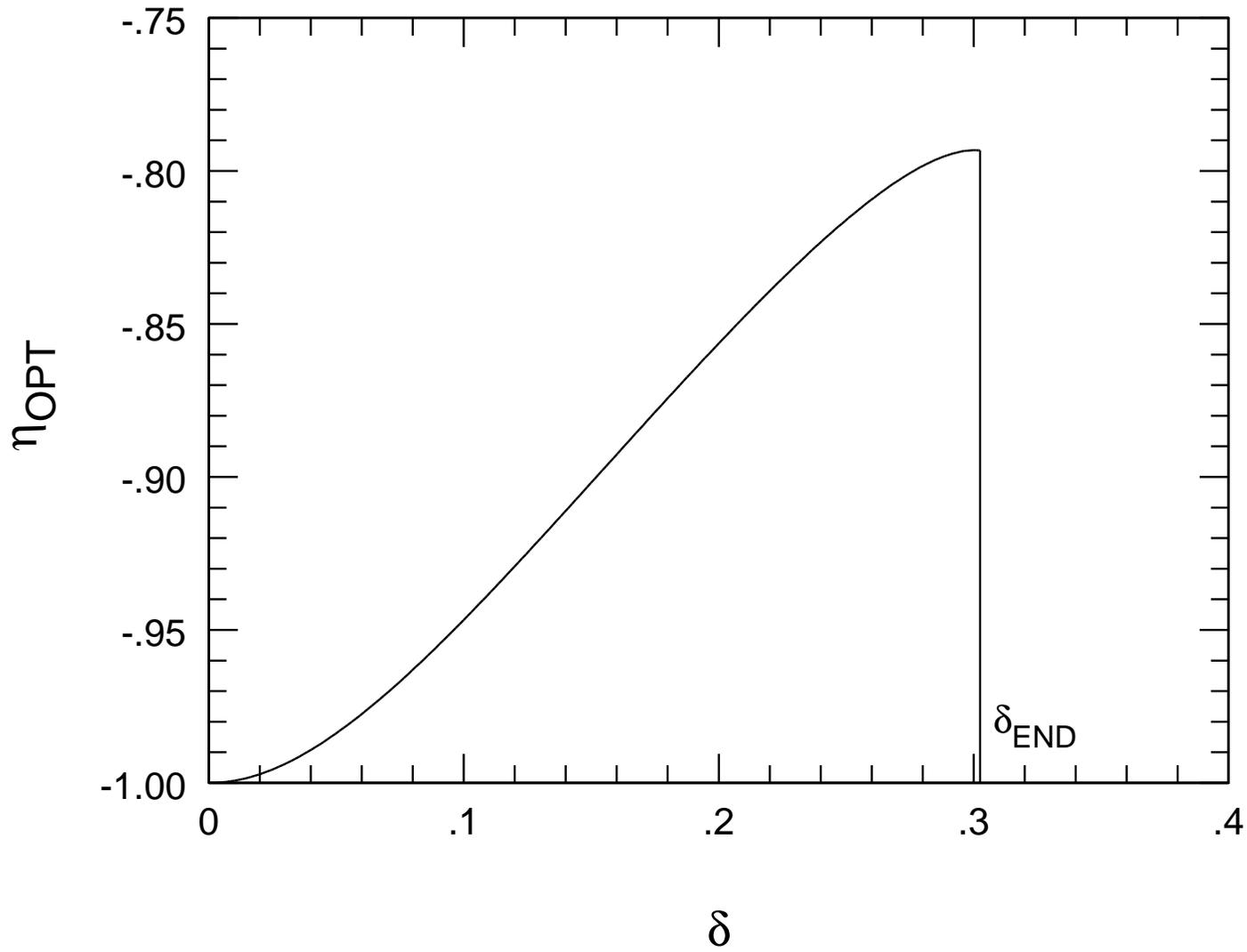

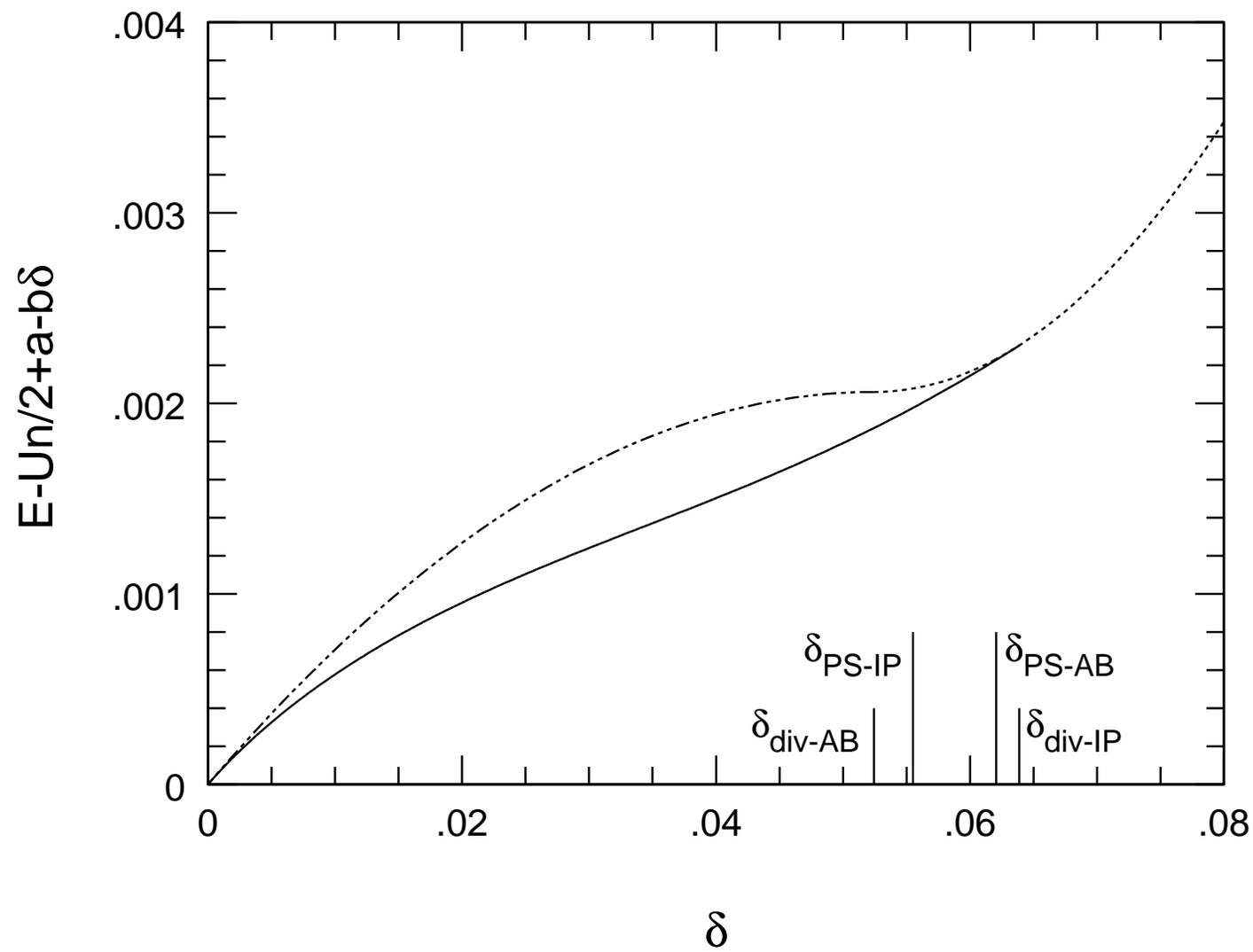

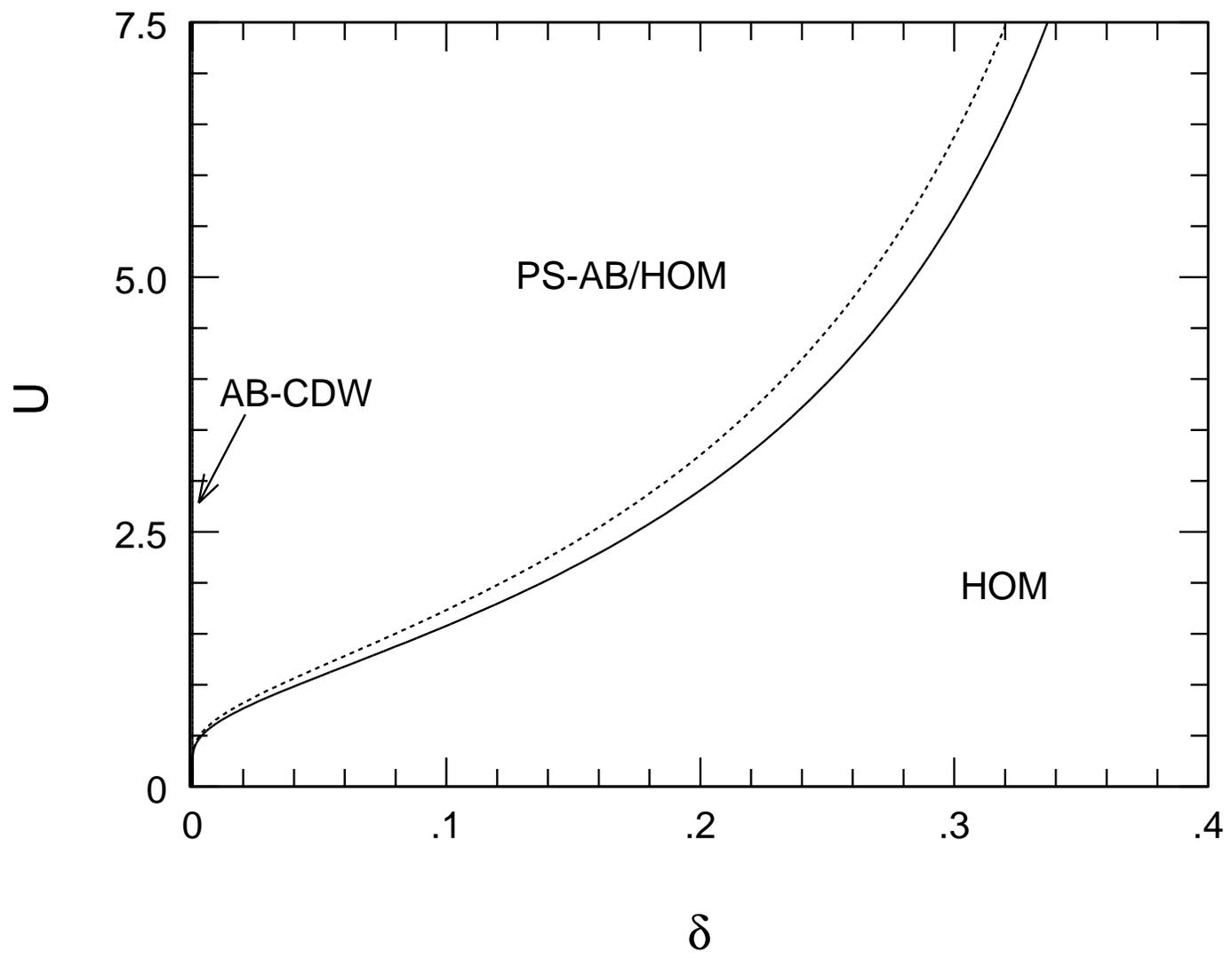

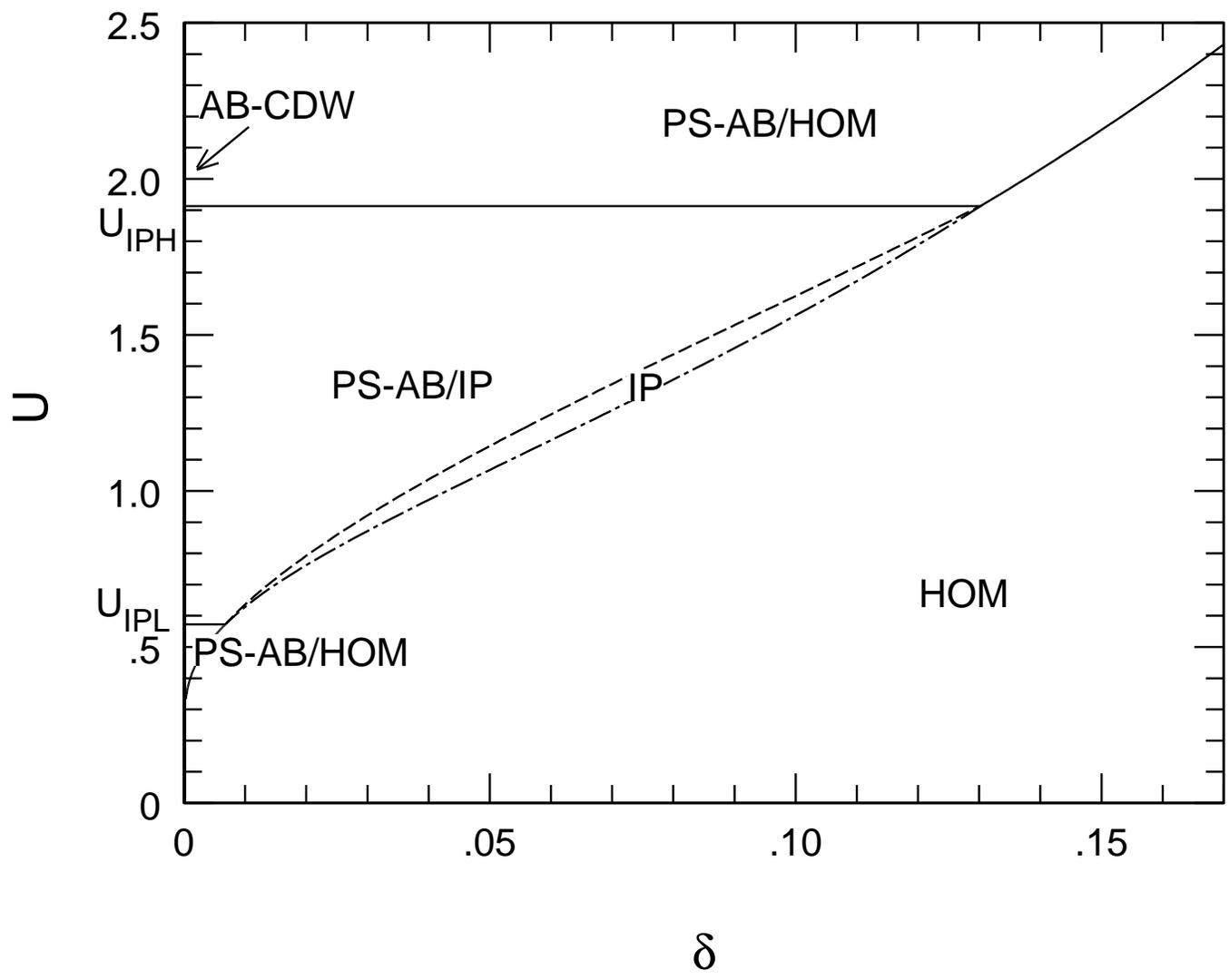

(a)

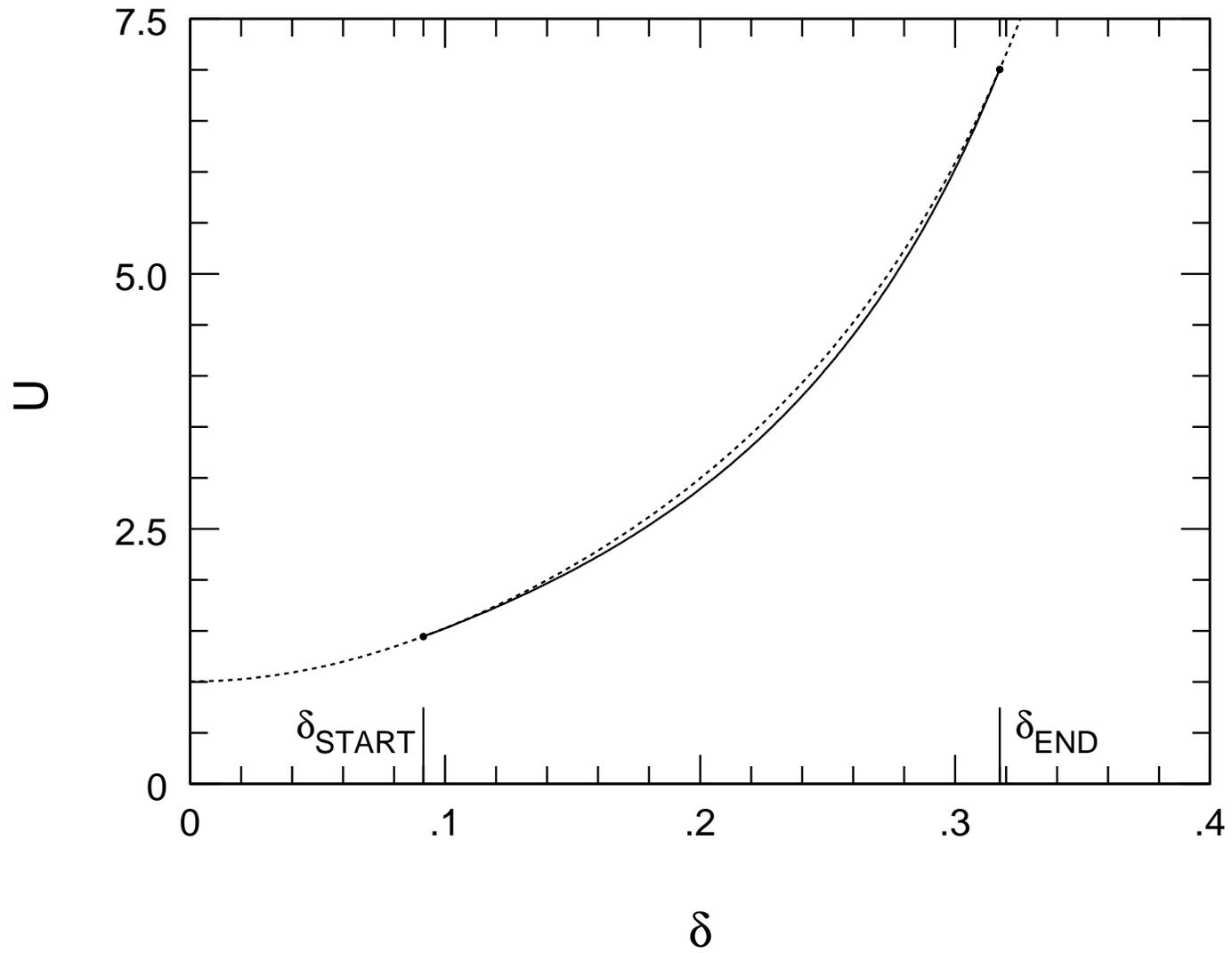

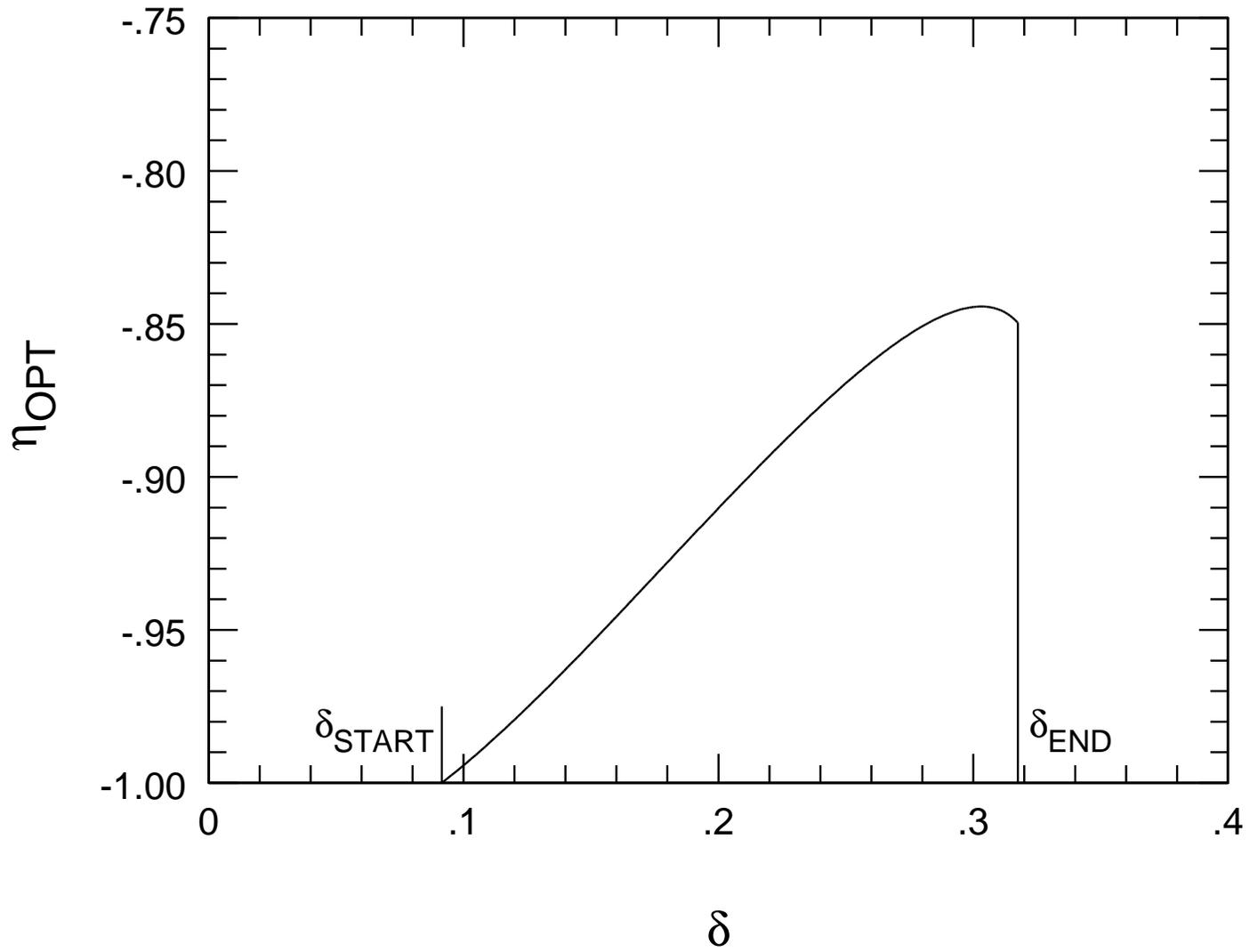

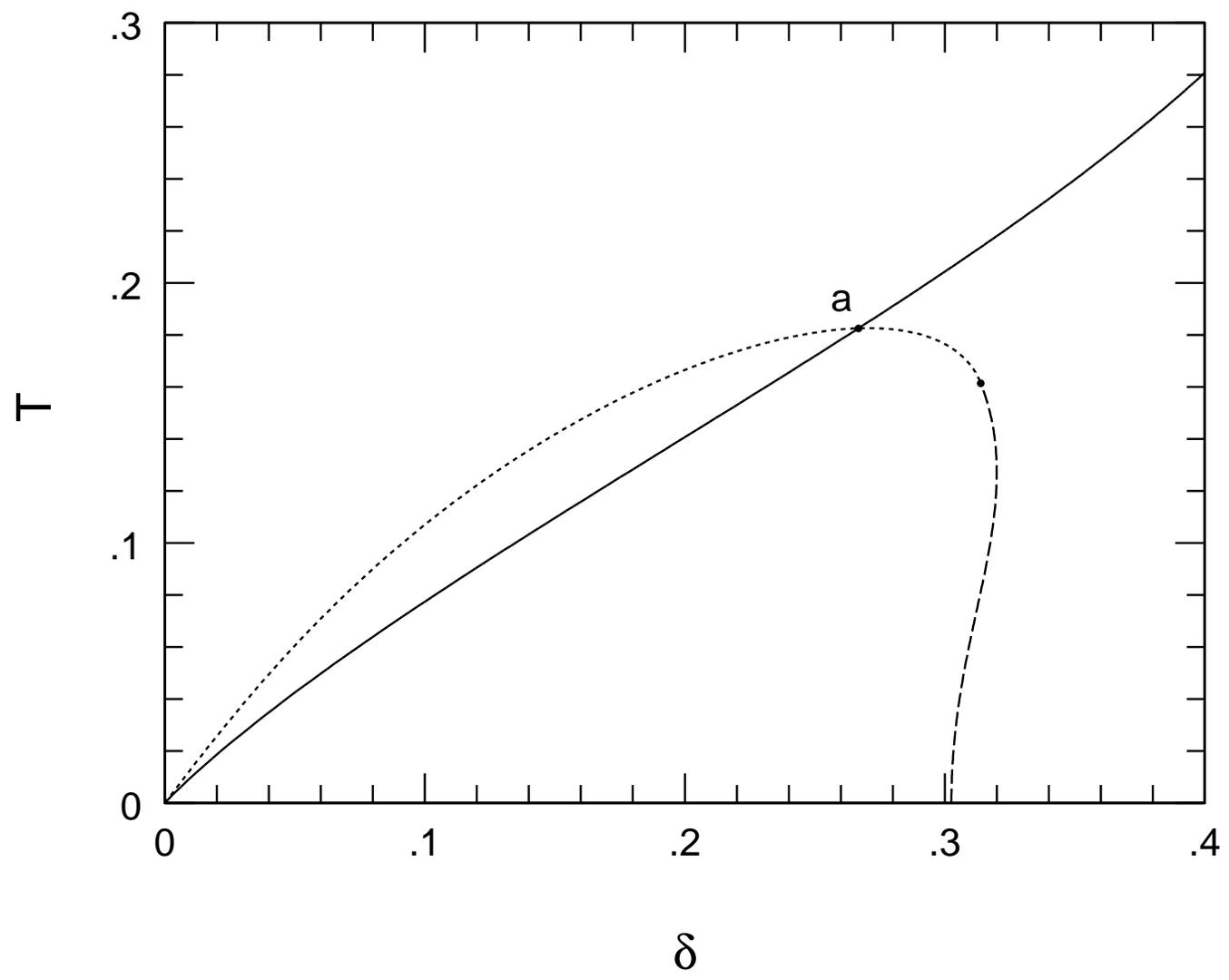

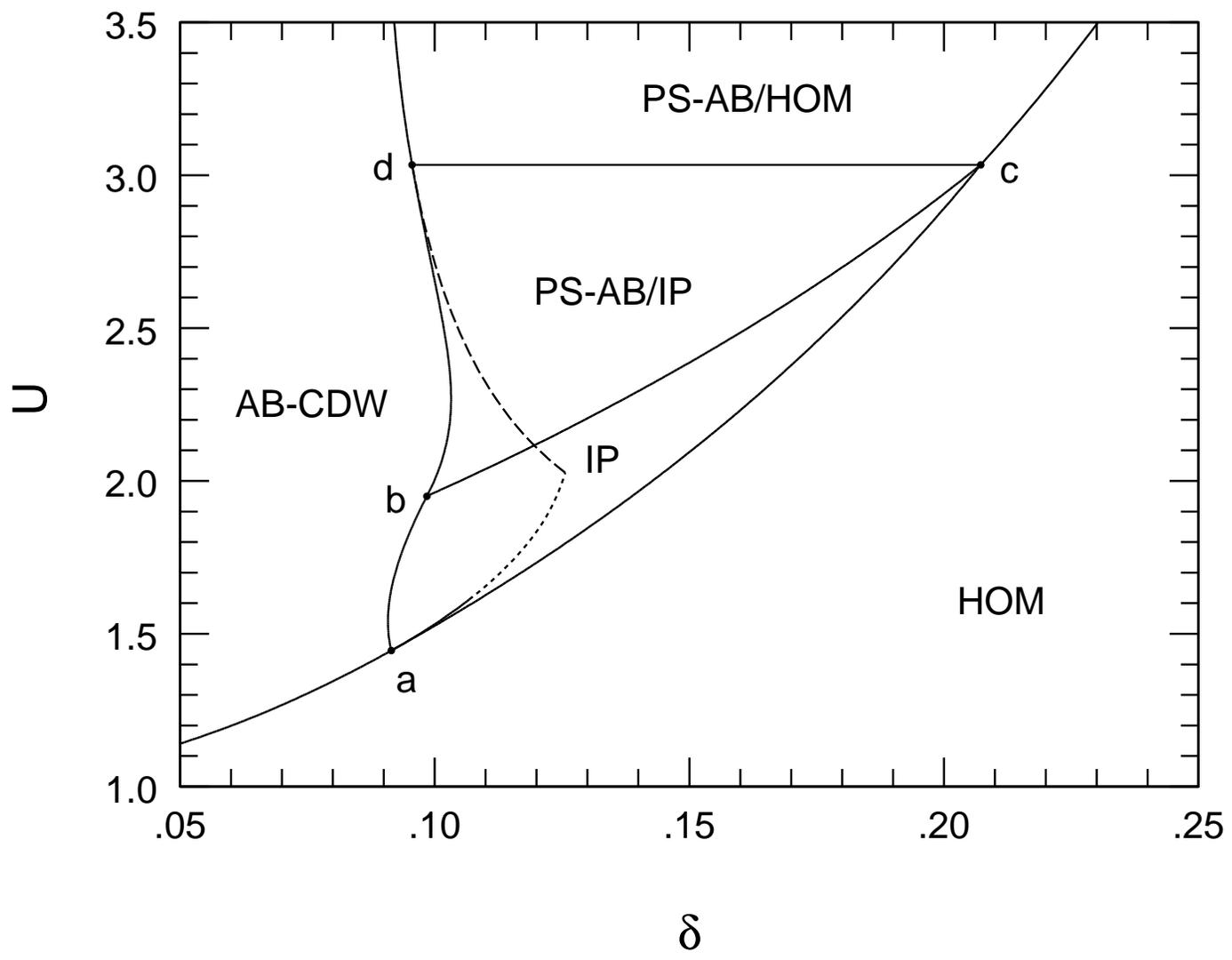

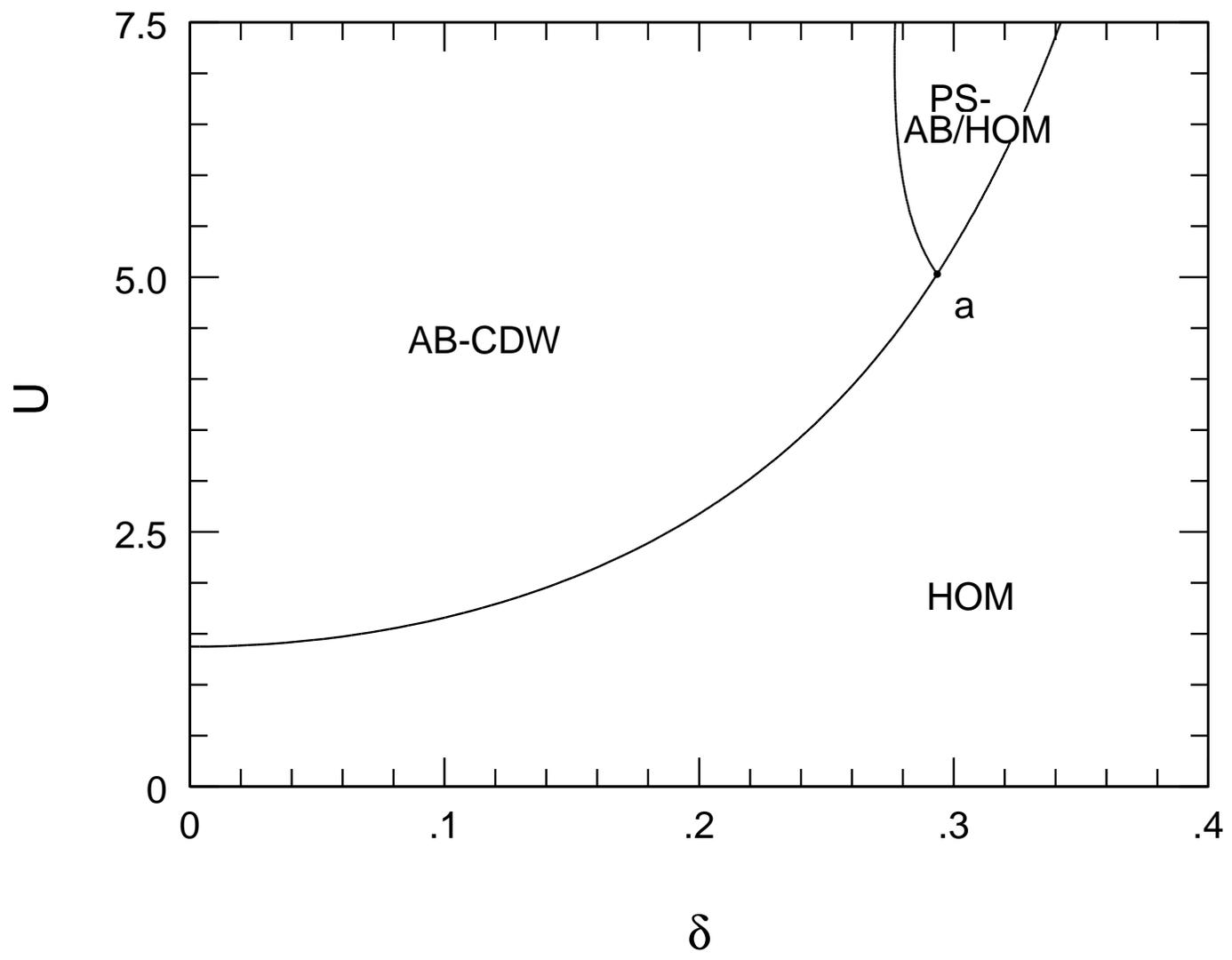

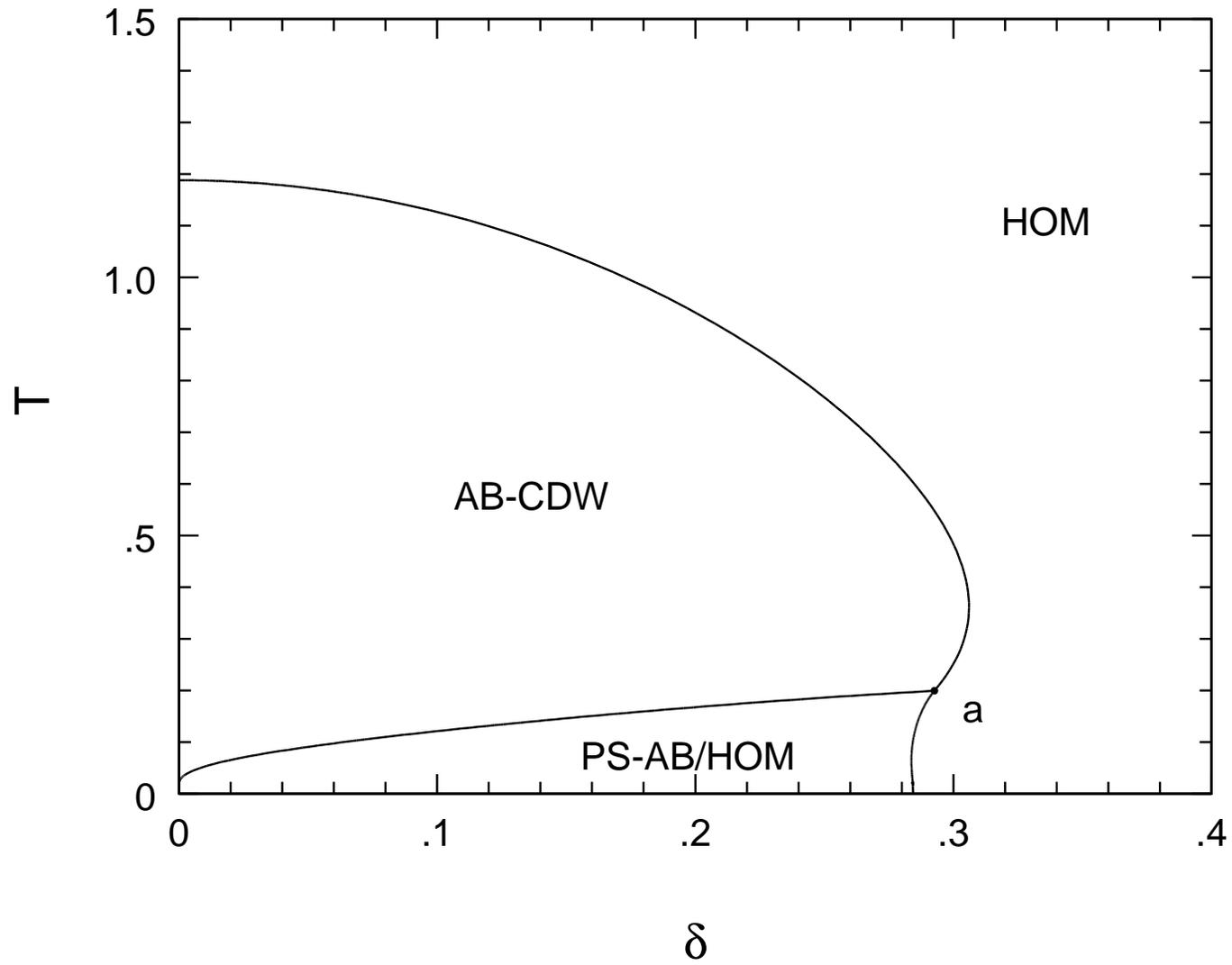

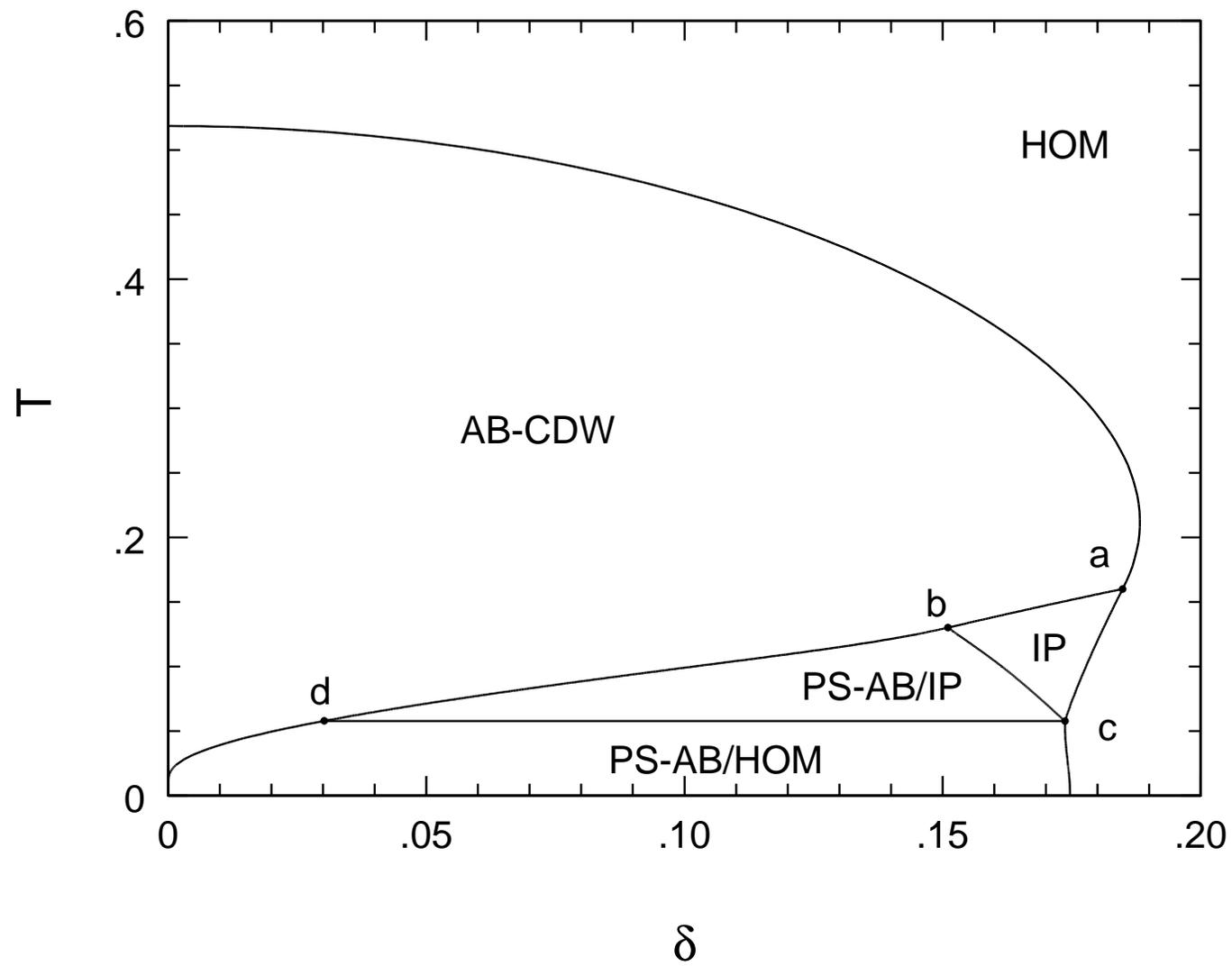

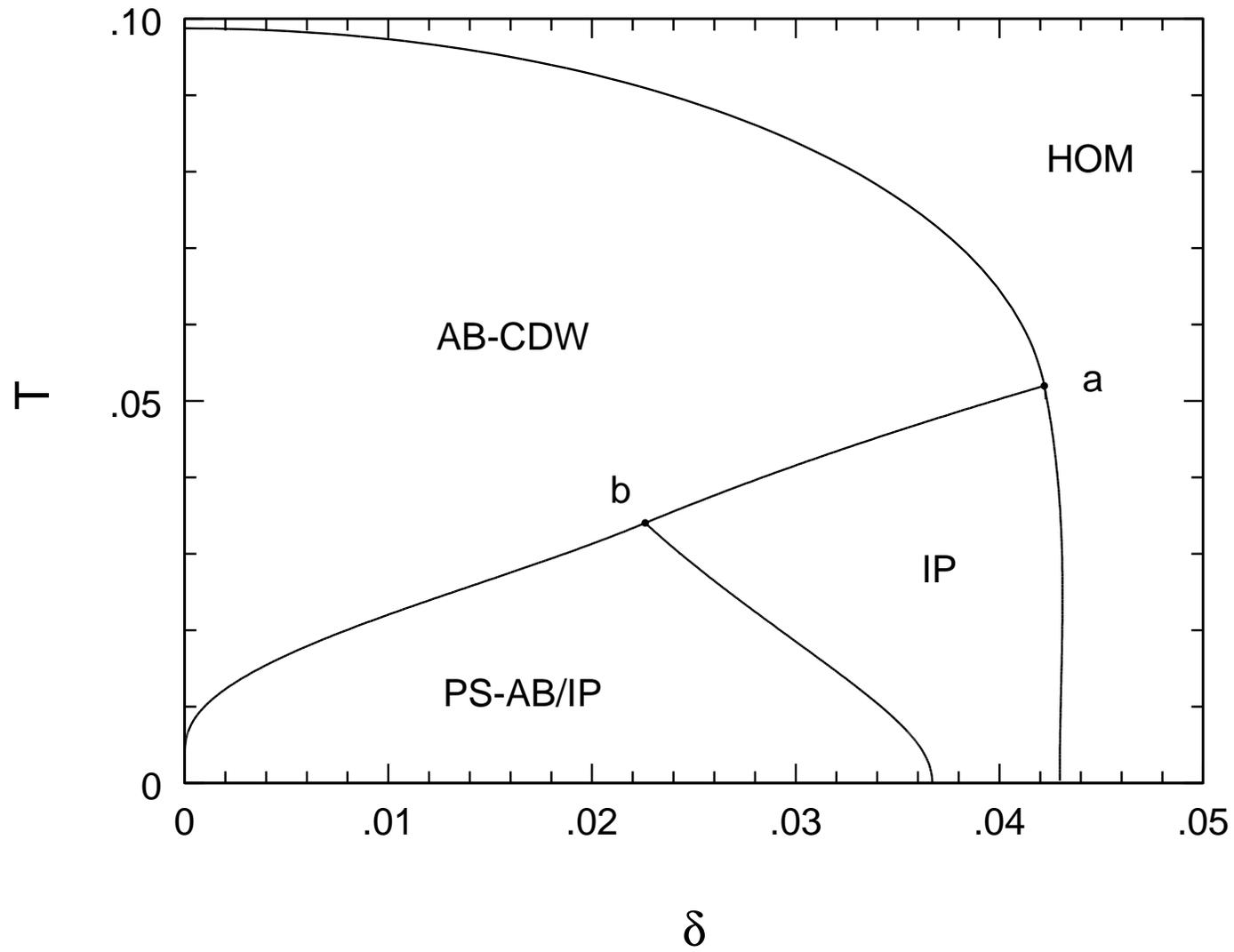

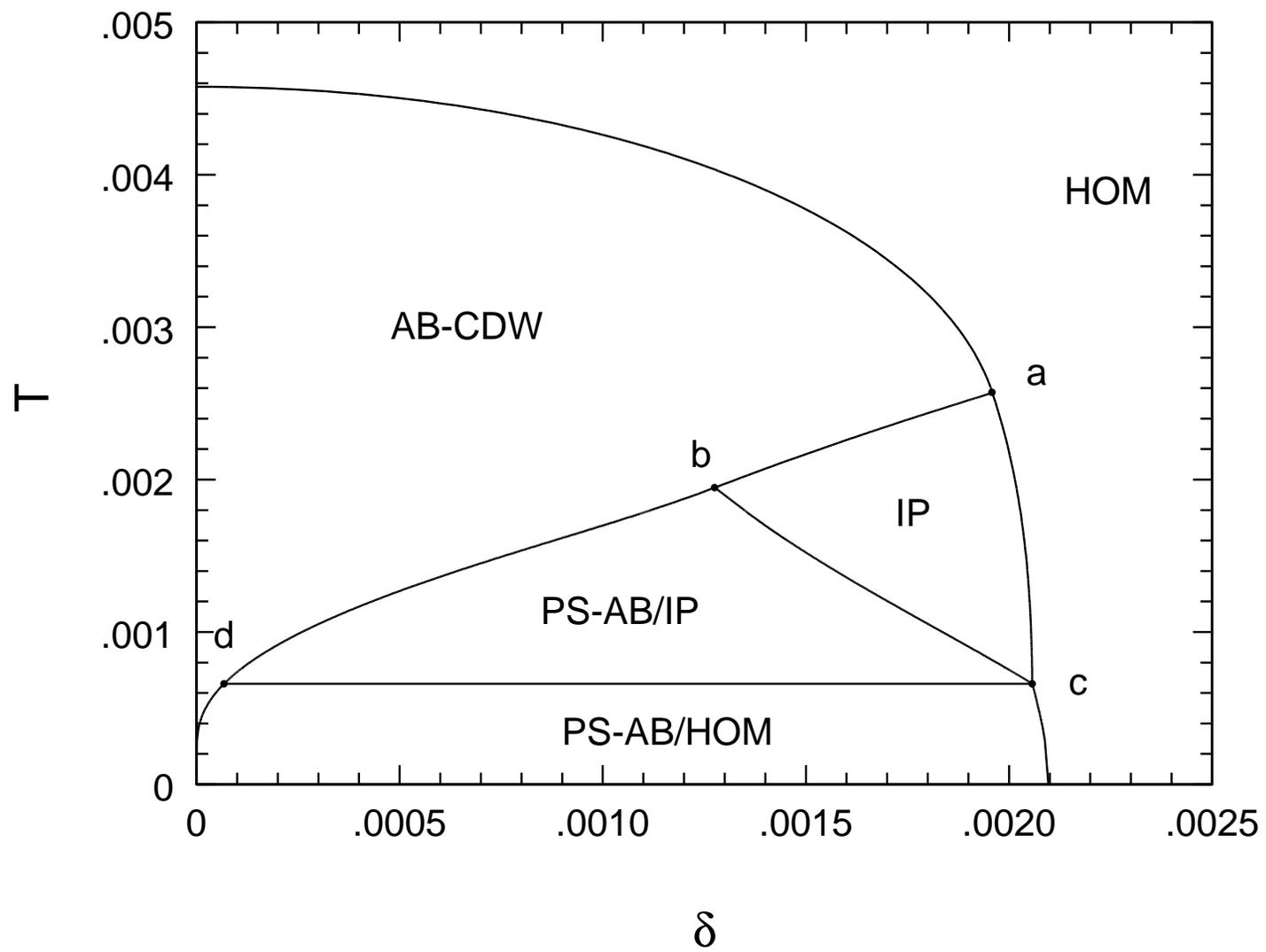